\documentclass[mathleft
]{an}

\usepackage{graphicx}
\usepackage{txfonts}
\newcounter{subtable}
\usepackage{times}
\newcommand{\vsini}{$v \sin i$}

\newcommand{\geneva}{{\sc geneva}}

\newcommand{\logg}{$\log g$}

\newcommand{\teff}{$T_{\mathrm{eff}}$}

\newcommand{\logl}{$\log(L/L_{\odot})$}

\begin{document}

% The following seven commands are intended for editorial usage and should be ignored by
% the author(s).
\Pagespan{789}{}% Document's page range. 
% If second parameter is left empty, the last page is computed automatically.
\Yearpublication{2006}%
\Yearsubmission{2005}%
\Month{11}%   
\Volume{999}%  
\Issue{88}% 
% \DOI{This.is/not.aDOI}% 

\title{New magnetic field measurements 
of $\beta$\,Cephei stars and Slowly Pulsating B stars\thanks{
Based on observations obtained at the European Southern Observatory, Paranal, Chile
(ESO programmes 078.D-0140(A), 078.D-0330(A), 079.D-0241(A), and 080.D-0383(A)).
}}

\author{{S. Hubrig\inst{1}\fnmsep\thanks{Corresponding author: \email{shubrig@eso.org}}}
%Example 
%for footnote, note the usage of the \texttt{fnmsep}
%command as separator between institute number and footnote mark} 
\and M. Briquet\inst{2}
\and P. De Cat\inst{3}
\and M. Sch\"oller\inst{4}
\and T. Morel\inst{5}
\and I. Ilyin\inst{6}
}
\titlerunning{Magnetic fields in $\beta$\,Cephei and SPB stars}
\authorrunning{S. Hubrig et al.}
\institute{
European Southern Observatory, Casilla 19001, Santiago 19, Chile
\and
Instituut voor Sterrenkunde, Katholieke Universiteit Leuven, Celestijnenlaan 200B, B-3001 Leuven, Belgium
\and
Koninklijke Sterrenwacht van Belgi\"e, Ringlaan 3, B-1180 Brussel, Belgium
\and
European Southern Observatory, Karl-Schwarzschild-Str. 2, D-85748 Garching bei M\"unchen, Germany
\and
Institut d'Astrophysique et de G\'eophysique, Universit\'e de Li\`ege, All\'ee du 6 Ao\^ut, B\^at. B5c, 4000 Li\`ege, Belgium
\and
Astrophysikalisches Institut Potsdam, An der Sternwarte 16, 14482 Potsdam, Germany 
}

%\received{}
%\accepted{}
%\publonline{later}


\abstract{
%We present the results of the continuation of our magnetic survey of a sample of B-type stars consisting of $\beta$\,Cephei stars,  Slowly Pulsating B (hereafter SPB) stars, candidate $\beta$\,Cephei and SPB stars, and a small number of normal B-type stars with FORS\,1 at the VLT.
%THIERRY
We present the results of the continuation of our magnetic survey with FORS\,1 at the VLT of a sample of 
B-type stars consisting of confirmed or candidate $\beta$\,Cephei stars and Slowly Pulsating B (hereafter SPB) 
stars, along with a small number of normal B-type stars.
%%%%%%%%%%%%
%of 16 confirmed and suspected $\beta$\,Cephei stars, 
%37 confirmed and suspected Slowly Pulsating B (SPB) stars and seven early to mid normal 
%B-type stars with FORS\,1 at the VLT. 
A weak mean longitudinal magnetic field of the order of a few hundred Gauss
was detected in three $\beta$\,Cephei stars and two stars suspected to be $\beta$\,Cephei stars, 
in five SPB stars and eight stars suspected to be SPB stars. Additionally, a longitudinal magnetic field 
at a level larger than 3$\sigma$ has been
diagnosed in two normal B-type stars, the nitrogen-rich early B-type star HD\,52089 and in the 
B5 IV star HD\,153716.
Roughly one third of $\beta$\,Cephei stars have detected magnetic fields: Out of 13 $\beta$\,Cephei stars 
studied to date with FORS\,1, four stars possess weak magnetic fields, 
and out of the sample of six suspected $\beta$\,Cephei stars two show a weak magnetic field. 
The fraction of magnetic SPBs and candidate SPBs is found to be higher: roughly half of the 34 SPB 
stars have been found to be magnetic and among the 16 candidate SPBs eight stars possess magnetic fields.
%THIERRY
%We are claiming that there is a higher proportion of magnetic stars in SPBs than in beta Cephei (53% vs 31%), but are we sure this is statistically significant? After all, we are still dealing with rather small statistics.
%We only have very few (and often a single) measurements for many stars. I think it is important to emphasize that these figures (53% and 31%) are lower limits.
%%%%%%%%%%%%
In an attempt to understand why only a fraction of pulsating stars exhibit magnetic fields, we
studied the position of magnetic and non-magnetic pulsating stars in the H-R diagram. 
We find that their domains in the H-R diagram largely overlap, and 
no clear picture emerges as to the possible evolution of the magnetic field across the main sequence.
It is possible that stronger fields tend to be found in stars with lower pulsating frequencies and 
smaller pulsating amplitudes. A somewhat similar trend is found if we consider a correlation
between the field strength and the \vsini{}-values, i.e.\
stronger magnetic fields tend to be found in more slowly rotating stars.
%Hints of some loose relations between the magnetic field strength, the dominant frequency and the corresponding 
%amplitude are found, though additional magnetic studies are needed to confirm the trends.
%It is also possible that the magnetic field is stronger in more slowly rotating stars.
%THIERRY
%I'm really not sure that such trends (even 'loose') are seen in the data. Instead of putting the emphasis on these rather inconclusive results, I'd rather emphasize in the abstract that a very significant fraction of pulsating B stars are magnetic. This is the most important result of your study, I think.
%%%%%%%%%%%%
%Recent results that a number of detected magnetic pulsating stars show chemical peculiarities such as 
%nitrogen excess and boron 
%deficiency (Morel et al.\ \cite{Morel2006}; Morel et al.\ \cite{Morel2008}) open a new perspective to use chemical anomalies as an indicator
%for target selection in  magnetic field surveys of pulsating B-type stars. 
}

\maketitle

\section{Introduction}

We started our systematic search for magnetic fields in pulsating B-type stars after the detection
of a weak magnetic field in two $\beta$~Cephei stars, in the prototype of the class, $\beta$~Cep itself, 
by Henrichs et al.\ (\cite{Henrichs2000}) and in V2052\,Oph by Neiner et al.\ (\cite{Neiner2003a}).
The first detection of a weak magnetic field in the SPB star $\zeta$~Cas
was reported by Neiner et al.\ (\cite{Neiner2003b}).
In our first publication on the magnetic survey of pulsating B-type stars (Hubrig et al.\ \cite{Hubrig2006}), 
we announced detections of weak mean longitudinal 
magnetic fields of the order of a few hundred~Gauss in 13~SPB stars and in 
the $\beta$~Cephei star $\xi^1$~CMa. Among the three $\beta$\,Cephei stars with detected magnetic 
fields, $\xi^1$\,CMa showed the largest mean
longitudinal field of the order of 300\,G.

However, the role of magnetic fields in modeling oscillations of B-type stars remains to be studied. 
Our previous search for correlations between the strength of the magnetic field and stellar fundamental
parameters, using available, rather scarce data, was unsuccessful.
%Our previous search for correlations between the variation of the magnetic field with the pulsation period
%using available, rather scarce data was unsuccessful. 
%THIERRY
%The reader (like me in this case...) may not understand why one of our primary goals is to establish a variation of the magnetic field strength along the pulsation period. I mean, is it something expected?
%%%%%%%%%%%%
We also did not find any hint of relations between the magnetic 
field strength and other stellar parameters. The position of the magnetic pulsating stars 
in the H-R diagram did not indicate a noticeable difference in the evolutionary stage between the non-magnetic  
and magnetic pulsating stars.
On the other hand, the whole sample under previous study contained only 14 stars with detected 
magnetic fields. Clearly, to obtain better statistics it is necessary to increase the sample of targets 
with magnetic field 
measurements. The aim of the current study is to analyse $\beta$\,Cep and SPBs as a group, such as the distributions of 
their magnetic fields and their relation to stellar fundamental parameters, that is stellar mass, 
effective temperature, projected rotation velocity,  evolutionary state in terms of 
elapsed fraction of main-sequence lifetime, and pulsation period.
%THIERRY
%replace 'life' by 'lifetime' throughout the text
%%%%%%%%%%%%
Magnetic field measurements have been collected with FORS\,1 at the VLT in the last two years.
With this new dataset and the previous dataset presented by Hubrig et al.\ (\cite{Hubrig2006}), we 
now obtained at least one measurement for each of 
the currently confirmed 34 SPB stars visible from the Southern hemisphere and for
13 $\beta$\,Cephei stars.
The SPB-like variability of most of the studied SPBs
was discovered by the Hipparcos satellite (Waelkens et al.\ \cite{Waelkens1998}).
Their membership in the SPB class has been confirmed by long-term photometric and spectroscopic
monitoring projects undertaken by members of the Institute of Astronomy of the University of Leuven.
We refer to De Cat (\cite{DeCat2007b}) for a recent review 
%of SPB stars.
%THIERRY
on SPB stars.
%%%%%%%%%%%%

Here we present the results of 98 magnetic field measurements in a sample of 60 stars, which includes 
confirmed and suspected $\beta$\,Cephei and SPB stars and a small fraction of early to mid normal B-type stars.
We describe the derivation of the fundamental parameters of the stars in our sample and discuss the occurrence and 
strength of their magnetic fields in the context of their position in the H-R diagram.

\section{The sample of pulsating stars}
\label{sect:sample}

Besides the confirmed SPB stars studied by long-term photometric and spectroscopic
monitoring projects, we enlarged our magnetic field search to several other Hipparcos SPB stars.
In the Hipparcos lightcurves, these targets show periods of the order of days, which correspond to the 
pulsation range of SPB stars. Furthermore, all these stars have been found to be located in the SPB 
instability strip.
Currently, we consider them as suspected SPB or candidate SPB
stars since additional studies of their variability are needed to definitely conclude on their nature.
Many chemically peculiar B-type magnetic stars, so-called Bp stars and ellipsoidal variables
are also found in the same part of the H-R diagram. For these stellar types 
%the observed variations are of the same order, but are usually attributed to rotation or binarity, instead of pulsations.
%THIERRY
the observed variations are operating on similar timescales, but are usually attributed to rotation or 
binarity, instead of pulsations (Briquet et al.\ \cite{Briquet2001}, \cite{Briquet2004}).
%%%%%%%%%%%%

The further goal of the 
%presented 
%THIERRY
present
%%%%%%%%%%%%
study was to enlarge the number of magnetic field measurements for a sample 
of $\beta$~Cephei stars.
Among them, the $\beta$~Cephei stars $\delta$~Cet and $\xi^1$~CMa were selected for monitoring 
because of the similarity of their pulsation behaviour. $\xi^1$~CMa was re-observed a couple of 
times with the aim to investigate its magnetic variability. As we mentioned in our 
previous work (Hubrig et al.\ \cite{Hubrig2006}),  $\delta$~Cet  and three $\beta$~Cephei stars with 
detected magnetic fields, V2052\,Oph, $\xi^1$\,CMa and
$\beta$\,Cep,  share common properties: all four are nitrogen enriched (Morel et al.\ \cite{Morel2006}), and all of them 
are either radial pulsators ($\xi^1$\,CMa) or their multiperiodic pulsations are dominated
by a radial mode ($\delta$\,Cet, $\beta$\,Cep, and V2052\,Oph).  The presence of
a magnetic field in these stars might play an important role to explain these
physical characteristics.
To search for the presence of possible differences in fundamental parameters between pulsating and non-pulsating 
stars we also selected seven non-pulsating normal early to mid B-type stars including the nitrogen-rich B 
star HD\,52089 recently studied by Morel et al.\ (\cite{Morel2008}).
%THIERRY
%As it is, it looks as if HD 52089 is N-rich, but the others are not. However, I've checked the literature and there are no abundance data for them, so we don't know.
%I seem to understand that we are using the stars in Table 3 as a control sample of non-pulsating stars, but I doubt that any of them have carefully been looked at for pulsations. Have they? Perhaps they are low-amplitude variables. The perfect control sample would be bona fide non pulsating stars like the rejected or poor beta Cephei candidates in Table 3 of Stankov & Handler (2005).
%%%%%%%%%%%%

\renewcommand{\thetable}{\arabic{table}\alph{subtable}}
\addtocounter{table}{0}
\setcounter{subtable}{1}

\begin{table*}
\caption{
The observed $\beta$\,Cephei stars. In the first three column we list the HD number, another 
identifier, the spectral type retrieved from the SIMBAD database and membership in a spectroscopic binary system.
An asterisk in front of the HD number denotes candidate $\beta$\,Cephei stars.
The effective temperature \teff{} and the surface gravity \logg{} are listed in Cols.~4 and 5 (see text).
%Observations in the Geneva photometric system are available for all targets.
%Mean Geneva magnitudes 
%were used to obtain the effective temperature {\bf \teff{}} and the surface 
%gravity \logg{} (listed in Cols.~4 and 5) with the method described in De Cat (2002). 
%The \teff{} and \logg{} values of 
%HD\,44743, HD\,46328, HD\,50707, HD\,129557 and HD\,180642 are inaccurate since an extrapolation 
%outside the calibration grid was needed. Other stellar parameters were derived from a grid of main-sequence 
%models calculated with the Code Li\'egeois d'\'Evolution Stellaire (version 18.2, written by R.~Scuflaire) 
%described as 'grid 2' in De Cat et al.\,(2006).
In Cols.~6 to 10 we present the stellar mass $M$, the radius $R$, the luminosity \logl{}
and the age of the star expressed as a fraction of its total main-sequence lifetime $f$. The last column gives \vsini{}-values.
%Fundamental parameters for the objects in our sample.
%In the first two columns we give the HD number and another identifier.
%In the following six columns we list spectral type, the logarithm of
%the effective temperature, the logarithm of the surface gravity, mass,
%stellar radius, and the logarithm of the stellar luminosity.
%The final two columns give the fraction of the main sequence lifetime
%for each individual star and its \vsini{}.
%{\change Uncertain parameter values are given in {\it italics}.}
}
\label{table1a}
\begin{center}
%\begin{tabular}{rrrr@{$\pm$}rr@{$\pm$}rr@{$\pm$}rr@{$\pm$}rr@{$\pm$}rr@{$\pm$}rr@{$\pm$}r}
\begin{tabular}{rccr@{$\pm$}lr@{$\pm$}lr@{$\pm$}lr@{$\pm$}lr@{$\pm$}lr@{$\pm$}lr@{$\pm$}l}
\hline
\multicolumn{1}{c}{HD} &
\multicolumn{1}{c}{Other} &
\multicolumn{1}{c}{Spectral} &
\multicolumn{2}{c}{\teff{}} &
\multicolumn{2}{c}{\logg{}} &
\multicolumn{2}{c}{$M/M_\odot$} &
%\multicolumn{2}{c}{age} &
\multicolumn{2}{c}{$R/R_\odot$} &
\multicolumn{2}{c}{\logl{}} &
%\multicolumn{2}{c}{Xc} &
\multicolumn{2}{c}{$f$} &
\multicolumn{2}{c}{\vsini}\\
\multicolumn{1}{c}{} &
\multicolumn{1}{c}{Identifier} &
\multicolumn{1}{c}{Type} &
\multicolumn{2}{c}{[$10^3$\,K]} &
\multicolumn{2}{c}{} &
\multicolumn{2}{c}{} &
%\multicolumn{2}{c}{[$10^6$\,yr]} &
\multicolumn{2}{c}{} &
\multicolumn{2}{c}{} &
%\multicolumn{2}{c}{} &
\multicolumn{2}{c}{[\%]} &
\multicolumn{2}{c}{[km~s$^{-1}$]}\\
\hline
       16582   & $\delta$~Cet & B2~IV     & 21.9 & 1.0 & 4.05 & 0.20 &  8.4 &  0.7  & 4.6 & 0.8 & 3.6 & 0.2 & 58 & 21 &  7  & 4   \\%16582
       29248   & $\nu$~Eri      & B2~III    & 23.0 & 1.1 & 3.92 & 0.20 & 10.1 &  0.8  & 6.1 & 0.9 & 4.0 & 0.1 & 79 & 13 & 21 & 12 \\%29248        
       %44743   & $\beta$~CMa   & B1~II-III & 26.4 & 1.2 & 3.79 & 0.20 & 13.5 &  1.0  & 7.51 & 1.1 & 4.4 & 0.1 & 82 &  9 & 11 &  7 \\%44743  
       44743   & $\beta$~CMa   & B1~II-III & {\it 26.4} & {\it 1.2} & {\it 3.79} & {\it 0.20} & 13.5 &  1.0  & 7.5 & 1.1 & 4.4 & 0.1 & 82 &  9 & 11 &  7 \\%44743  
       46328   & $\xi^1$~CMa   & B1~III    & {\it 27.1} & {\it 1.2} & {\it 3.83} & {\it 0.20} & 13.7 &  0.9  & 7.1 & 0.9 & 4.4 & 0.1 & 76 & 10 & 20 &  7 \\%46328  
       50707   & 15~CMa        & B1~Ib     & {\it 26.1} & {\it 1.2} & {\it 3.89} & {\it 0.20} & 12.8 &  1.2  & 6.8 & 1.2 & 4.3 & 0.2 & 75 & 13 & 20& 12    \\%50707  $\ast$
$\ast$ 55958   & GG~CMa        & B2~IV     & 19.2 & 0.9 & 4.15 & 0.20 &  6.4 &  0.4 & 3.6 & 0.5 & 3.2 & 0.1 & 45 & 23 & \multicolumn{2}{c}{---}    \\%55958
       61068   & PT~Pup        & B2~II     & 23.8 & 1.1 & 4.01 & 0.20 & 10.1 &  0.8  & 5.4 & 0.9 & 3.9 & 0.2 & 64 & 18 & 10   &  9  \\%61068        
$\ast$ 74575   & $\alpha$~Pyx  & B1.5~III  & 23.8 & 1.1 & 3.89 & 0.20 & 10.7 &  0.9  & 6.3 & 1.0 & 4.0 & 0.1 & 79 & 12 & 11 & 2   \\%74575        
      111123   & $\beta$~Cru   & B0.5~III, SB1 & 27.3 & 1.3 & 3.73 & 0.20 & 14.1 &  0.7  & 7.5 & 0.7 & 4.4 & 0.1 & 81 & 6 & 16 & 9 \\%111123       
      129557   & BU~Cir        & B2~III    & {\it 24.9} & {\it 1.1} & {\it 4.06} & {\it 0.20} & 10.5 &  0.7  & 5.0 & 0.8 & 3.9 & 0.1 & 51 & 21 &   53 & 16   \\%129557 $\ast$
      129929   & V836~Cen      & B3~V      & 23.9 & 1.1 & 4.03 & 0.20 & 10.1 &  0.7  & 5.2 & 0.8 & 3.9 & 0.1 & 61 & 19 &  8 &  5 \\%129929       
$\ast$ 132200  & $\kappa$~Cen  & B2~IV     & 19.8 & 0.9 & 4.02 & 0.20 &  7.2 &  0.5  & 4.4 & 0.7 & 3.4 & 0.2 & 64 & 19 & 11   & 7   \\%132200       
$\ast$ 136504  & $\epsilon$~Lup & B2~IV-V, SB   & 19.3 & 0.9 & 3.89 & 0.20 &  7.4 &  0.6 & 5.2 & 0.9 & 3.5 & 0.2 & 83 & 13 & 41   & 9  \\%136504       
$\ast$ 171034  & HR~6960       & B2~IV-V   & 18.8 & 0.9 & 3.81 & 0.20 &  7.3 &  0.5 & 5.3 & 0.7 & 3.5 & 0.1 & 87 &  9 &  107  & 16 \\%171034       
$\ast$ 172910  & HR~7029       & B2.5~V    & 18.8 & 0.9 & 4.23 & 0.20 &  6.1 &  0.4 & 3.4 & 0.4 & 3.1 & 0.1 & 37 & 20 &  7  &  4  \\%172910       
       180642  & V1449~Aql     & B1.5~II-III & {\it 27.8} & {\it 1.3} & {\it 4.14} & {\it 0.20} & 12.6 &  1.0 & 5.1 & 0.8 & 4.1 & 0.1 & 36 & 22 & 15  & 9  \\%180642 $\ast$
\hline
\end{tabular}
\end{center}
\end{table*}

\addtocounter{table}{-1}
\addtocounter{subtable}{1}

\begin{table*}
\caption{
Same as in Table~\ref{table1a} for the observed SPB stars. An asterisk denotes candidate SPB stars.
%Observations in the Geneva photometric system are available for all targets. Mean Geneva magnitudes 
%were used to obtain the effective temperature log(Teff) and the surface 
%gravity log(g) with the method described in De Cat (2002). The log(Teff) and log(g) values of 
%HD\,46328, HD\,50707, HD\,52089 and HD\,180642 are inaccurate because an extrapolation 
%outside the calibration grid was needed. Other stellar parameters were derived from a grid of main-sequence 
%models calculated with the Code Li\'egeois d' \'Evolution Stellaire (version 18.2, Scuflaire et al.\ \cite{Scuflaire2008}) 
%described as 'grid 2' in De Cat et al.\,(2006). The mass $M$, the radius $R$, the luminosity log(L/L) 
%and the age of the star expressed as a fraction of its total main-sequence lifetime $f$ are also presented.  
%Fundamental parameters for the objects in our sample.
%In the first two columns we give the HD number and another identifier.
%In the following six columns we list spectral type, the logarithm of
%the effective temperature, the logarithm of the surface gravity, mass,
%stellar radius, and the logarithm of the stellar luminosity.
%The final two columns give the fraction of the main sequence lifetime
%for each individual star and its \vsini{}.
%{\change Uncertain parameter values are given in {\it italics}.}
}
\label{table1b}
\begin{center}
\begin{tabular}{rccr@{$\pm$}lr@{$\pm$}lr@{$\pm$}lr@{$\pm$}lr@{$\pm$}lr@{$\pm$}lr@{$\pm$}l}
\hline
\multicolumn{1}{c}{HD} &
\multicolumn{1}{c}{Other} &
\multicolumn{1}{c}{Spectral} &
\multicolumn{2}{c}{\teff{}} &
\multicolumn{2}{c}{\logg{}} &
\multicolumn{2}{c}{$M/M_\odot$} &
%\multicolumn{2}{c}{age} &
\multicolumn{2}{c}{$R/R_\odot$} &
\multicolumn{2}{c}{\logl{}} &
%\multicolumn{2}{c}{Xc} &
\multicolumn{2}{c}{$f$} &
\multicolumn{2}{c}{\vsini}\\
\multicolumn{1}{c}{} &
\multicolumn{1}{c}{Identifier} &
\multicolumn{1}{c}{Type} &
\multicolumn{2}{c}{[$10^3$\,K]} &
\multicolumn{2}{c}{} &
\multicolumn{2}{c}{} &
%\multicolumn{2}{c}{[$10^6$\,yr]} &
\multicolumn{2}{c}{} &
\multicolumn{2}{c}{} &
%\multicolumn{2}{c}{} &
\multicolumn{2}{c}{[\%]} &
\multicolumn{2}{c}{[km~s$^{-1}$]}\\
\hline
         3379 & 53~Psc     & B2.5I~V   & 17.3 & 0.8 & 4.16 & 0.20 &  5.4 &  0.4 & 3.3 & 0.5 & 2.9 & 0.1 & 45 & 23 & 33 & 17 \\%3379         
$\ast$  11462 & CG~Hyi     & B8~V      & 12.6 & 0.6 & 4.31 & 0.20 &  3.2 &  0.2 & 2.2 & 0.2 & 2.0 & 0.1 & 31 & 17 &23  & 13    \\%11462                
$\ast$  23958 & HR~1186    & B8~V      & 12.5 & 0.6 & 4.05 & 0.20 &  3.5 &  0.3 & 3.0 & 0.5 & 2.3 & 0.2 & 65 & 18 & 320   & 16 \\%23958    
        24587 & 33~Eri     & B5~V, SB1 & 13.9 & 0.6 & 4.26 & 0.20 &  3.7 &  0.2 & 2.5 & 0.3 & 2.3 & 0.1 & 36 & 19 & 28 &  1 \\%24587        
        25558 & 40~Tau     & B3~V      & 16.4 & 0.8 & 4.22 & 0.20 &  4.9 &  0.3 & 3.0 & 0.4 & 2.8 & 0.1 & 39 & 21 & 14 &  8 \\%25558        
        26326 & GU~Eri     & B5~IV     & 15.2 & 0.7 & 4.14 & 0.20 &  4.4 &  0.3 & 3.0 & 0.5 & 2.7 & 0.1 & 49 & 22 & 11 &  6 \\%26326        
$\ast$  26739 & GY~Eri     & B5~IV     & 15.3 & 0.7 & 4.09 & 0.20 &  4.6 &  0.3 & 3.3 & 0.5 & 2.7 & 0.1 & 57 & 21 &16  &8   \\%26739        
$\ast$  27742 & V1141~Tau  & B8~IV-V   & 12.7 & 0.6 & 4.13 & 0.20 &  3.4 &  0.2 & 2.7 & 0.4 & 2.2 & 0.1 & 53 & 21 & 175  & 24  \\%27742        
        28114 & V1143~Tau  & B6~IV     & 14.6 & 0.7 & 4.00 & 0.20 &  4.5 &  0.3 & 3.5 & 0.6 & 2.7 & 0.2 & 72 & 17 &  9 &  5 \\%28114        
        28475 & V1144~Tau  & B5~V      & 15.1 & 0.7 & 3.97 & 0.20 &  4.8 &  0.4 & 3.8 & 0.6 & 2.8 & 0.2 & 76 & 15 & 15 & 8   \\%28475        
$\ast$  29376 & V1148~Tau  & B3~V, SB  & 16.5 & 0.8 & 4.19 & 0.20 &  5.0 &  0.3 & 3.1 & 0.4 & 2.8 & 0.1 & 42 & 22 & 89  & 15   \\%29376        
$\ast$  33331 & TU~Pic     & B5~III    & 13.0 & 0.6 & 4.22 & 0.20 &  3.4 &  0.2 & 2.5 & 0.3 & 2.2 & 0.1 & 40 & 21 & 27 & 14  \\%33331        
        34798 & YZ~Lep     & B5~IV-V, SB ? & 15.6 & 0.7 & 4.25 & 0.20 &  4.5 &  0.3 & 2.8 & 0.3 & 2.6 & 0.1 & 37 & 20 & 34 &  2 \\%34798        
        37151 & V1179~Ori  & B8~V      & 12.9 & 0.6 & 4.34 & 0.20 &  3.3 &  0.2 & 2.2 & 0.2 & 2.1 & 0.1 & 27 & 15 & 11  & 6    \\%37151        
        39844 & $\epsilon$~Dor & B6~V  & 13.7 & 0.6 & 3.89 & 0.20 &  4.3 &  0.3 & 3.8 & 0.6 & 2.7 & 0.1 & 84 & 11 &  9  & 5   \\%39844               
$\ast$  40494 & $\gamma$~Col & B2.5~IV & 15.9 & 0.7 & 3.72 & 0.20 &  5.7 &  0.3 & 4.8 & 0.4 & 3.1 & 0.1 & 93 &  5 &  96  & 16  \\%40494 
        45284 & BD-07~1424 & B8, SB2   & 14.7 & 0.7 & 4.40 & 0.20 &  3.9 &  0.2 & 2.4 & 0.2 & 2.4 & 0.1 & 19 & 11 & 71 &  6 \\%45284        
        53921 & V450~Car   & B9~IV, SB2 & 13.7 & 0.6 & 4.23 & 0.20 &  3.7 &  0.2 & 2.6 & 0.3 & 2.3 & 0.1 & 39 & 21 & 17 & 10 \\%53921        
$\ast$  55718 & V363~Pup   & B3~V      & 16.1 & 0.7 & 4.17 & 0.20 &  4.8 &  0.3 & 3.1 & 0.5 & 2.8 & 0.1 & 45 & 23 &179  & 4   \\%55718                
        69144 & NO~Vel     & B2.5~IV   & 15.9 & 0.7 & 3.80 & 0.20 &  5.5 &  0.3 & 4.6 & 0.5 & 3.1 & 0.1 & 89 &  8 &  67 & 2   \\%69144        
        74195 & o~Vel      & B3~IV     & 16.2 & 0.7 & 3.91 & 0.20 &  5.5 &  0.4 & 4.3 & 0.7 & 3.0 & 0.2 & 82 & 13 &  9 &  5 \\%74195        
        74560 & HY~Vel     & B3~IV, SB1 & 16.2 & 0.7 & 4.15 & 0.20 &  4.9 &  0.3 & 3.1 & 0.5 & 2.8 & 0.1 & 46 & 24 & 13 &  7 \\%74560        
        85953 & V335~Vel   & B2~III    & 18.4 & 0.8 & 3.91 & 0.20 &  6.8 &  0.6 & 4.9 & 0.8 & 3.4 & 0.2 & 82 & 13 & 18 & 10 \\%85953        
$\ast$ 128585 & IS~Lup     & B3~IV     & 16.6 & 0.8 & 3.89 & 0.20 &  5.8 &  0.4 & 4.5 & 0.7 & 3.1 & 0.1 & 83 & 12 &132  & 12  \\%128585       
       140873 & 25~Ser     & B8~III, SB2 & 13.9 & 0.6 & 4.35 & 0.20 &  3.7 &  0.2 & 2.4 & 0.2 & 2.3 & 0.1 & 26 & 15 & 70 &  2 \\%140873       
$\ast$ 152511 & V847~Ara   & B5~III    & 14.8 & 0.7 & 4.23 & 0.20 &  4.2 &  0.3 & 2.7 & 0.4 & 2.5 & 0.1 & 39 & 21 &26  & 15   \\%152511       
$\ast$ 152635 & V1070~Sco  & B7~II     & 13.8 & 0.6 & 4.31 & 0.20 &  3.6 &  0.2 & 2.4 & 0.3 & 2.3 & 0.1 & 30 & 17 & 5  & 3   \\%152635       
$\ast$ 163254 & V1092~Sco  & B2~IV/V   & 18.1 & 0.8 & 4.10 & 0.20 &  6.0 &  0.4 & 3.7 & 0.6 & 3.1 & 0.1 & 53 & 22 & 110 & 6   \\%163254       
$\ast$ 169467 & $\alpha$~Tel & B3~IV   & 16.7 & 0.8 & 4.12 & 0.20 &  5.2 &  0.4 & 3.3 & 0.5 & 2.9 & 0.1 & 51 & 22 & 14 & 8   \\%169467   
       169820 & BD+14~3533 & B9~V      & 11.8 & 0.5 & 4.26 & 0.20 &  2.9 &  0.2 & 2.2 & 0.3 & 1.9 & 0.1 & 37 & 20 & 136  & 7   \\%169820       
       179588 & V338~Sge   & B9~IV     & 12.2 & 0.6 & 4.28 & 0.20 &  3.0 &  0.2 & 2.2 & 0.3 & 2.0 & 0.1 & 34 & 19 & 20 & 12   \\%179588       
       181558 & V4199~Sgr  & B5~III    & 14.7 & 0.7 & 4.16 & 0.20 &  4.2 &  0.3 & 2.9 & 0.5 & 2.5 & 0.1 & 46 & 23 &  6 &  4 \\%181558       
$\ast$ 183133 & V4372~Sgr  & B2~IV     & 16.7 & 0.8 & 3.99 & 0.20 &  5.5 &  0.4 & 4.0 & 0.7 & 3.0 & 0.2 & 72 & 17 & 25  & 13   \\%183133       
       191295 & V1473~Aql  & B7~III    & 13.1 & 0.6 & 4.14 & 0.20 &  3.6 &  0.2 & 2.7 & 0.4 & 2.3 & 0.1 & 51 & 22 & 9 & 6   \\%191295       
$\ast$ 205879 & DK~Oct     & B8~V      & 12.5 & 0.6 & 4.23 & 0.20 &  3.2 &  0.2 & 2.4 & 0.3 & 2.1 & 0.1 & 40 & 21 & 11 & 6   \\%205879       
       206540 & BD+10~4604 & B5~IV     & 14.0 & 0.6 & 4.11 & 0.20 &  4.0 &  0.3 & 3.0 & 0.5 & 2.5 & 0.1 & 55 & 21 &  8 &  5 \\%206540       
$\ast$ 215573 & $\xi$~Oct  & B6~IV     & 14.0 & 0.6 & 4.09 & 0.20 &  4.0 &  0.3 & 3.0 & 0.5 & 2.5 & 0.1 & 58 & 20 &  5 &  2 \\%215573       

\hline
\end{tabular}
\end{center}
\end{table*}

\addtocounter{table}{-1}
\addtocounter{subtable}{1}

\begin{table*}
\caption{
Same as in Table~\ref{table1a} for the observed normal B stars and the N-rich star HD~52089.
%The \teff{} and \logg{} values of 
%HD\,52089 are inaccurate since an extrapolation 
%outside the calibration grid was needed.
%Fundamental parameters for the objects in our sample.
%In the first two columns we give the HD number and another identifier.
%In the following six columns we list spectral type, the logarithm of
%the effective temperature, the logarithm of the surface gravity, mass,
%stellar radius, and the logarithm of the stellar luminosity.
%The final two columns give the fraction of the main sequence lifetime
%for each individual star and its \vsini{}.
%{\change Uncertain parameter values are given in {\it italics}.}
}
\label{table1c}
\begin{center}
\begin{tabular}{rccr@{$\pm$}rr@{$\pm$}rr@{$\pm$}rr@{$\pm$}rr@{$\pm$}lr@{$\pm$}rr@{$\pm$}r}
\hline
\multicolumn{1}{c}{HD} &
\multicolumn{1}{c}{Other} &
\multicolumn{1}{c}{Spectral} &
\multicolumn{2}{c}{\teff{}} &
\multicolumn{2}{c}{\logg{}} &
\multicolumn{2}{c}{$M/M_\odot$} &
%\multicolumn{2}{c}{age} &
\multicolumn{2}{c}{$R/R_\odot$} &
\multicolumn{2}{c}{\logl{}} &
%\multicolumn{2}{c}{Xc} &
\multicolumn{2}{c}{$f$} &
\multicolumn{2}{c}{\vsini}\\
\multicolumn{1}{c}{} &
\multicolumn{1}{c}{Identifier} &
\multicolumn{1}{c}{Type} &
\multicolumn{2}{c}{[$10^3$\,K]} &
\multicolumn{2}{c}{} &
\multicolumn{2}{c}{} &
%\multicolumn{2}{c}{[$10^6$\,yr]} &
\multicolumn{2}{c}{} &
\multicolumn{2}{c}{} &
%\multicolumn{2}{c}{} &
\multicolumn{2}{c}{[\%]} &
\multicolumn{2}{c}{[km~s$^{-1}$]}\\
\hline 
24626   & i~Eri      & B6~V    & 14.1 & 0.6 & 4.24 & 0.20 &  3.8 &  0.2 & 2.6 & 0.4 & 2.4 & 0.1 & 38 & 20 & 29 & 2   \\%24626
52089   & 21~CMa     & B2~Iab  & {\it 25.1} & {\it 1.2} & {\it 3.82} & {\it 0.20} & 12.5 &  1.3  & 7.3 & 1.3 & 4.3 & 0.2 & 84 & 10 &  28  & 2   \\%52089
142378  & 47~Lib     & B2/B3~V & 16.1 & 0.7 & 4.38 & 0.20 &  4.5 &  0.2 & 2.6 & 0.2 & 2.6 & 0.1 & 21 & 12 &230  & 24    \\%142378       
153716  & HR~6320    & B5~IV   & 15.5 & 0.7 & 4.07 & 0.20 &  4.7 &  0.3 & 3.4 & 0.5 & 2.8 & 0.1 & 60 & 20 & 187  & 10   \\%153716       
164245  & HR~6708    & B7~IV   & 12.9 & 0.6 & 4.03 & 0.20 &  3.7 &  0.3 & 3.1 & 0.5 & 2.4 & 0.2 & 68 & 17 &  \multicolumn{2}{c}{---}\\%164245       
166197  & NSV~10304  & B1~V    & 23.8 & 1.1 & 3.75 & 0.20 & 11.5 &  1.1 & 7.2 & 1.1 & 4.2 & 0.1 & 88 &  7 & 229  & 15   \\%166197       
169033  & BD-12~5024 & B5~V    & 12.0 & 0.6 & 3.82 & 0.20 &  3.6 &  0.2 & 3.6 & 0.4 & 2.4 & 0.1 & 88 &  8 & 160  & 24   \\%169033       
\hline
\end{tabular}
\end{center}
\end{table*}
%THIERRY
%For the stars I have analyzed, I'm not surprised to see differences between Teff, logg, etc., as the methods are completely different, but I notice a poor agreement for the vsini values (e.g. 16 vs 48 km/s for beta Cru!). I'm not sure why that is. 
%Do you think it would be useful to put the nitrogen over carbon abundance ratios ([N/C]) in these tables, so that the reader can judge how N-rich the stars are? Except the magnetic stars we already know about, there are values for 3 stars without field detection in the literature (HD 61068, HD 129557 and HD 215573).
%%%%%%%%%%%%

All stars are very bright ($V\le8$) and their pulsational behaviour has 
been intensively studied during the last years. 
The fundamental parameters of 
the studied $\beta$\,Cephei stars (both confirmed and suspected) are presented in Table~\ref{table1a}, those 
of SPB and candidate SPB stars in Table~\ref{table1b}, and  Table~\ref{table1c} contains the 
fundamental parameters of normal B-type stars and the N-rich star HD\,52089.

%THIERRY
%I suppose the paper by Stankov & Handler (2005) has been used to decide whether a star is a confirmed or candidate beta Cephei?
%I would rename Tables 1-3 in Tables 1a-c.
%%%%%%%%%%%%
For all the objects in our sample, observations in the \geneva{} photometric
system are available.
Mean \geneva{} magnitudes were used to obtain the effective temperature
\teff{} and the surface gravity \logg{} with the method described by
K\"unzli et al.\ (\cite{Kunzli1997})
(Cols.~4 and 5 in Tables~\ref{table1a}--\ref{table1c}).
%(assuming solar composition;
%The errors on \teff{} and \logg{} were estimated to be 0.020~dex and
%0.20~dex, respectively, for every star.
%The mass $M$, the radius $R$, the luminosity \logl, the hydrogen abundance
%in the core \xc, and the fraction of the total main-sequence lifetime
%\ms{} was determined as the mean values
In Table~\ref{table1a} the \teff{} and \logg{} values of 
HD\,44743, $\xi^1$\,CMa, HD\,50707, HD\,129557, and HD\,180642 are inaccurate because an extrapolation 
outside the calibration grid was needed for their determination.
The same extrapolation was used for HD\,52089 in Table~\ref{table1c}.
All these values are set in {\it italics}.
To derive other stellar parameters, a grid of main-sequence models has been used,
which was calculated with the Code Li\'egeois d'\'Evolution Stellaire
(version 18.2, Scuflaire et al.\ \cite{Scuflaire2008}), assuming solar composition.
For a detailed description see ``grid 2'' in De Cat et al.\ (\cite{DeCat2006}).
The mass $M$, the radius $R$, the luminosity \logl, and the age of 
the star expressed as a fraction of its total main-sequence lifetime
$f$ are presented 
in Cols.~6 to 9 in Tables~\ref{table1a}--\ref{table1c}.
We note that the SIMBAD spectral classification as giants or supergiants  of a
number of $\beta$~Cephei and SPB stars is in all cases misleading, probably due to the low $v \sin i$ values of
the considered stars. The fundamental parameters presented in Tables~\ref{table1a}--\ref{table1c}
clearly show that only one star in our
sample, HD\,40494, is in an advanced evolutionary state, just behind the terminal-age main sequence (TAMS). 
%correspond to the average values and their standard deviation derived
%from the selected models.
%If a significant part of the error box falls outside the covered part
%of the main-sequence, the values are given in {\it italics}.
%%These values should be treated with caution.
%For the most massive stars of our sample, HD\,44743, HD\,46005, and HD\,46328, an extrapolation of the grid 
%models was needed and thus their parameters are less reliable. The star HD\,45284
%falls outside the main sequence. The uncertain parameters for these four stars are given in italics
%in Table~\ref{table1}.
%We point out that four stars fall outside the main sequence: HD\,45284, HD\,46005, HD\,140873 and HD\,160124, so 
%that the stellar parameters are less reliable for these stars. The upper value of 15 $M_\odot$ of the grid models 
%is insufficient to fully cover the observed error box in \teff{} and \logg{} of HD\,44743, HD\,46328, and 
%HD\,111123. Moreover the \teff{} and \logg{} of HD\,44743, HD\,46005 and HD\,46328 are less accurate because an 
%extrapolation was needed for their determination. All these less reliable parameter values are given in italic 
%in Table~\ref{table1}. In addition we point out that the physical parameters of binary systems (indicated as SB1 
%and SB2 in Table~\ref{table1}), especially the multiple-lined ones, also have to be treated with caution.

For almost all studied stars numerous spectroscopic observations have been obtained in
previous years at the ESO La Silla observatory and 
Observatoire de Haute-Provence
(Aerts et al.\ \cite{Aerts1998}; Mathias et al.\ \cite{Mathias2001}; Uytterhoeven et al.\ \cite{Uytterhoeven2001}; De Cat \& Aerts \cite{DeCat2002};
Aerts et al.\ \cite{Aerts2004a}, \cite{Aerts2004b}).
To estimate the projected rotational velocity 
\vsini{} (Col.~10 in Tables~\ref{table1a}--\ref{table1c}),
an average of all spectra has been used for single stars.
For binaries with a well-known orbit, the orbital motion was removed before averaging the spectra. 
In case of double-lined systems, 
usually the spectrum with the maximum observed separation between the components was chosen 
for the \vsini{} determination.
%\vsini{} - values have been found in the literature. They are presented in Tables???
%without an accuracy estimate}.
For slowly rotating $\beta$\,Cephei and SPB stars we selected several unblended absorption lines:
the $\lambda$4560\,\AA{} Si\,III-triplet
and/or the $\lambda$4130\,\AA{} Si\,II-doublet.
For rapidly 
rotating stars, only the $\lambda$4481\,\AA{} Mg\,I-line has been used.
We applied the method of least squares fitting
with rotationally broadened synthetic profiles using a Gaussian intrinsic
width but without taking into account pulsational broadening. 
%(hereafter ``fitting method'') and the Fourier method (\cite{Gray1992}).

For fifteen stars we do not have high-resolution spectroscopic observations.
%for a small fraction of pulsating stars. 
For thirteen stars it was possible to gather \vsini{} values 
from the literature.
The \vsini{} values for
HD\,27742, HD\,40494, HD\,61068, HD\,129557, HD\,136504, HD\,142378, HD\,169033, and HD\,171034 were taken from 
Abt et al.\ (\cite{Abt2002}).
For HD\,23958 and HD\,169820 the \vsini{} values were found in Royer et al.\ (\cite{Royer2002}).
For the normal B-type star HD\,24626 
we used the \vsini{} value from Hempel \& Holweger (\cite{HempelHolweger2003}) and the data from 
Balona (\cite{Balona1975}) were used for the other two normal B-type stars, HD\,153716 and HD\,166197.

\section{Spectropolarimetric observations}

The spectropolarimetric observations have been carried out in the 
years 2006 to 2008 at the
European Southern Observatory with FORS\,1 (FOcal Reducer low dispersion
Spectrograph) mounted on the 8-m Melipal telescope of the VLT.
This multi-mode instrument is equipped with polarization analyzing optics
comprising super-achromatic half-wave and quarter-wave phase retarder plates,
and a Wollaston prism with a beam divergence of 22$\arcsec$ in standard
resolution mode.
For the major part of observations we used the GRISM\,600B in the wavelength range 3480--5890\,\AA{}
to cover all hydrogen Balmer lines from H$\beta$ to the Balmer jump.
The observations at the end of August - beginning of September 2007 have been carried 
out  with  a new mosaic detector
with blue optimised E2V chips, which was implemented in FORS\,1 at the beginning of April 2007.
It has a  pixel size of 15\,$\mu$m (compared to 24\,$\mu$m for the
previous Tektronix chip) and higher efficiency
in the wavelength range below 6000\,\AA{}.
With the new mosaic detector
and the grism 600B we are 
%also able now 
%THIERRY
 now also able
%%%%%%%%%%%%
 to cover a much larger
spectral range, from 3250 to 6215\,\AA{}.
Two observations of $\delta$~Cet at the end of 2006 have been carried out 
with GRISM 600R in the wavelength range 5240--7380\,\AA{}. 
In all observations a slit width of 0$\farcs$4 was used to obtain a spectral resolving power 
of $R\sim2000$ with GRISM 600B and $R\sim3000$ with GRISM 600R.
During the observing run in August/September 2007 two $\beta$~Cep stars, $\xi^1$\,CMa and $\delta$~Cet, 
have been observed with the GRISM\,1200B covering the 
%H Balmer
%THIERRY
Balmer
%%%%%%%%%%%%
lines from H$\beta$ to H$8$, and a slit width
of 0$\farcs$4 to obtain a spectral resolving power of $R\sim4000$.

Usually, we took four to eight continuous series of two exposures for each star in 
our sample 
%using a standard readout mode with high gain (A,1$\times$1,high) and 
with the retarder waveplate oriented at two different angles, +45$^\circ$ and $-$45$^\circ$. 
%For the observations carried out in 2005 we used a non-standard readout mode with low 
%gain (A,1$\times$1,low), which allowed to increase the signal-to-noise ratio of 
%gain (A,1$\times$1,low), which provided a broader dynamic range, hence 
%allowed us to increase the signal-to-noise ratio of individual spectra 
%by a factor of $\approx$2 in comparison to the earlier observations.
All stars in our 
sample are bright and since the errors of the measurements of the 
polarization with FORS\,1 are determined by photon counting statistics, 
a signal-to-noise ratio 
of a few thousands can be reached for bright stars within $\sim$30\,min.
More details on the observing technique with FORS\,1 can be found elsewhere 
(e.g., Hubrig et al.\ \cite{Hubrig2004a}, \cite{Hubrig2004b}).
%Determination of the longitudinal magnetic field using the FORS\,1 spectra is achieved by 
%measuring the circular polarization of opposite sign induced in the wings of hydrogen Balmer lines, by the Zeeman effect. 
%Measurement of circular polarization in magnetically sensitive 
%lines is the most direct means of detecting magnetic fields on stellar surfaces. 
The mean longitudinal magnetic field is the average over the stellar hemisphere
visible at the time of observation of the component of the magnetic field
parallel to the line of sight, weighted by the local emergent spectral line
intensity.
It is diagnosed from the slope of a linear regression of $V/I$ versus the quantity
$-\frac{g_{\rm eff}e}{4\pi{}m_ec^2} \lambda^2 \frac{1}{I} \frac{{\mathrm d}I}{{\mathrm d}\lambda} \left<B_z\right> + V_0/I_0$,
where $V$ is the Stokes parameter which measures the circular polarization,
$I$ is the intensity observed in unpolarized light,
$g_{\rm eff}$ is the effective Land\'e factor,
$e$ is the electron charge,
$\lambda$  is the wavelength expressed in \AA{},
$m_e$ the electron mass,
$c$ the speed of light,
${{\rm d}I/{\rm d}\lambda}$ is the derivative of Stokes $I$,
and $\left<B_z\right>$ is the mean longitudinal field.
Our experience from a study of a large sample of magnetic and non-magnetic
Ap and Bp stars revealed that this regression technique is very robust
and that detections with $B_z > 3\sigma$ result only for stars
possessing magnetic fields.

To search for temporal variability of the magnetic field of $\xi^1$\,CMa we planned to obtain a series of
spectropolarimetric observations with the 2.56\,m Nordic Optical Telescope (NOT, La Palma) using the 
SOFIN echelle spectrograph.
However, due to bad weather conditions, only one rather noisy high resolution spectropolarimetric observation
with an exposure time of 30\,sec and a S/N ratio of $\sim$150 was obtained on 
September 13, 2008.
SOFIN is a high-resolution echelle spectrograph  mounted at the Cassegrain focus
of NOT (Tuominen et al.\ \cite{Tuominen1999}) and equipped with three optical cameras providing different resolving 
powers of 30\,000, 80\,000, and 160\,000.
$\xi^1$\,CMa was observed with the low-resolution camera with $R=\lambda/\Delta\lambda\approx30000$.
We used a 2K Loral CCD detector to register 40 echelle orders partially covering the range from 3500 to 10000\,\AA\ 
with a length of the spectral orders of about 140\,\AA\ at 5500\,\AA.
The polarimeter is located in front of the entrance slit of the spectrograph and consists of a fixed 
calcite beam splitter aligned along the slit and a rotating super-achromatic quarter-wave plate. Two spectra 
polarized in opposite sense are recorded simultaneously for each echelle order providing sufficient separation 
by the cross-dispersion prism below 7000\,\AA. Two such exposures with the quarter-wave plate angles separated 
by $90^\circ$ are necessary to derive circularly polarized spectra.
The spectra are usually reduced with the 4A software package (Ilyin \cite{Ilyin2000}). Bias subtraction, 
master flat-field correction, 
scattered light subtraction, and weighted extraction of spectral orders comprise the standard steps of the image 
processing. A ThAr spectral lamp is used for wavelength calibration, taken before and after each target exposure 
to minimize temporal variations in the spectrograph. 
%The Stokes I and V spectra are formed from the polarized spectra by taking the respective averages.  
%To search for temporal variability of the magnetic field of $\xi^1$\,CMa
%we planned to obtain a number spectropolarimetric observations
%with the Nordic Optical Telescope (NOT, La Palma) using the SOFIN echelle spectrograph.
%However, due to bad weather conditions, only a single rather noisy high resolution spectropolarimetric spectrum with 
%the exposure time of 30\,sec and a S/N ratio of $\sim$150 was recorded during our run on September 13, 2009
%SOFIN is a high-resolution echelle spectrograph, mounted at the Cassegrain focus of NOT 
%(Tuominen et al.\ \cite{Tuominen1999}).
%It is equipped with three cameras providing different resolving powers (WHICH ONES???).
%The spectrum of $\xi^1$\,CMa
%was obtained using the third camera, which allows to reach a resolution $R=\lambda/\Delta\lambda\approx30~000$.
%We used a $1152\times298$ pixel EEV CCD detector to register about 12 echelle orders, 
%partially covering the range from 4000 to 7000 ŽÃ…. The length of an individual order depends on the wavelength, but is 
%typically about 40\,\AA. 
%The spectra were reduced with the 4A software package (Ilyin \cite{Ilyin2000}).

%\input{table/table.tex}

\setcounter{subtable}{1}

\begin{table}
\caption{
The mean longitudinal magnetic field measurements for the 
$\beta$~Cephei stars in our sample, observed with FORS\,1.
In the first two columns we give the HD number
and the modified Julian date of the middle of the exposures.
The measured mean longitudinal magnetic fields $\left<B_{\mathrm z}\right>$
using all lines or only hydrogen lines are presented in Cols.~3 and 4. 
All quoted errors are 1$\sigma$ uncertainties.
In Col.~5 we identify new detections by ND and confirmed detections by CD.
We note that all claimed detections have a significance of at
least 3$\sigma$, determined from the formal uncertainties we derive. 
These measurements are indicated in bold face.
%we give the rms longitudinal magnetic field
%and the reduced $\chi^2$ for all measurements in columns~4 and 5.
}
\label{table2a}
\begin{center}
%\begin{tabular}{rcrp{2mm}rcr}
\begin{tabular}{rcr@{$\pm$}rr@{$\pm$}rc}
\hline
\multicolumn{1}{c}{\raisebox{2mm}{\rule{0mm}{2mm}}HD} &
\multicolumn{1}{c}{MJD} &
\multicolumn{2}{c}{$\left<B_{\mathrm z}\right>_{\rm all}$} &
\multicolumn{2}{c}{$\left<B_{\mathrm z}\right>_{\rm hydr}$} &
\multicolumn{1}{c}{Comment} \\
\multicolumn{1}{c}{} &
\multicolumn{1}{c}{} &
\multicolumn{2}{c}{[G]} &
\multicolumn{2}{c}{[G]} &
\multicolumn{1}{c}{} \\
\hline
%16582 & 54014.20764 & $-$112 & 36 & $-$142 & 52 &  \\
%16582 & 54109.06597& $-$88 & 24 & $-$86 & 39 &  \\
16582 & 54014.208 & $-$112 & 36 & $-$142 & 52 &  \\
      & 54109.066 & $-$88 & 24 & $-$86 & 39 &  \\
      & 54040.170 & 12 & 31 & 26 & 42 &  \\
      & 54047.129 &141  &83  &188  &106 &  \\
      & 54343.259 &$-$12  & 11 &$-$21  & 15  &  \\
      & 54344.200 & {\bf $-$40} & {\bf 12} & {\bf $-$75} & {\bf 14}& ND  \\
      & 54344.264 & {\bf $-$44} & {\bf 12} & {\bf $-$66} &{\bf 21}& ND   \\
      & 54345.203 &$-$42 &12  & $-$27 & 15 &  \\
      & 54345.245 &{\bf $-$49}  & {\bf 13} &{\bf $-$59} &{\bf 15} & ND  \\
      & 54345.293 &$-$31  & 11  &{\bf $-$44}  &{\bf 14} & ND  \\
29248 & 54086.286 & 63 & 43 & 66 & 55&  \\
44743 & 54046.360 & 122 & 52 & 198 & 72 &  \\
46328 & 54061.325 & {\bf 369} & {\bf 42} & {\bf 360} &{\bf  45} & CD \\
      & 54107.266 & {\bf 312} & {\bf 43} & {\bf 319} & {\bf 46}  & CD \\
      & 54114.028 & {\bf 309} & {\bf 35} & {\bf 347} & {\bf 38} & CD  \\
      & 54114.182 & {\bf 364} & {\bf 35} & {\bf 382} & {\bf 47}  & CD \\
      & 54116.108 & {\bf 307} & {\bf 45} & {\bf 276} & {\bf 58}  & CD \\
      & 54155.086 & {\bf 308} & {\bf 47} & {\bf 349} & {\bf 35}  & CD \\
      & 54343.371 & {\bf 345} & {\bf 11} & {\bf 379} & {\bf 15}  & CD \\
      & 54345.338 & {\bf 366} & {\bf 11} & {\bf 400} & {\bf 12}  & CD \\
      & 54345.414 & {\bf 340} & {\bf 11} & {\bf 378} & {\bf 18}  & CD \\
      & 54548.982 & {\bf 324} & {\bf 55} & {\bf 297} & {\bf 87}  & CD\\  
      & 54549.995 & {\bf 380} & {\bf 37} & {\bf 332} & {\bf 55}  & CD\\
50707 & 54107.318 & {\bf 163} & {\bf 52} & 157 & 58& ND \\
      & 54345.372 & {\bf 149} & {\bf 19} & {\bf 123} &{\bf 27}&ND \\ 
55958 & 54343.413 &$-$108  & 45 & $-$97 & 48 \\
61068 & 54107.338 &$-$12  & 41 &$-$14  &45&  \\
74575 & 54082.341 & 142  &48  &{\bf $-$219}  & {\bf 60}& ND \\
      & 54109.150 &132  &50  & {\bf 184} & {\bf 60}& ND \\
111123 & 54155.209 &107 & 74 & 45 & 86& \\ 
       & 54157.202& 95 & 41 & 86 & 50 & \\ 
129557 & 54158.229 &41  &26  & 77 & 28 \\
129929 &   54177.219 & 168 & 59 &147  & 67& \\ 
       & 54343.978 &$-$50  &33  &$-$24  & 41& \\
132200 & 54343.995 &135  & 61 &175  & 86&  \\	
136504 & 54344.998 &{\bf $-$156} & {\bf 34} &{\bf $-$128}& {\bf 36}& ND\\  
171034 & 54344.130 & $-$58 & 33 & $-$69 & 37 & \\
172910 & 54345.176 & 69& 36 & 72 & 43 &\\
180642 & 54343.159 &$-$55 & 33 &$-$93  & 38 &\\
       & 54344.084 &{\bf 166}  &{\bf 41}  & 160 & 55& ND \\
\hline
\end{tabular}
\end{center}
\end{table}

\addtocounter{table}{-1}
\addtocounter{subtable}{1}

\begin{table}
\caption{
Same as in Table~\ref{table2a} for the observed SPB stars.
}
\label{table2b}
\begin{center}
%\begin{tabular}{rcrp{2mm}rcr}
\begin{tabular}{rcr@{$\pm$}rr@{$\pm$}rc}
\hline
\multicolumn{1}{c}{\raisebox{2mm}{\rule{0mm}{2mm}}HD} &
\multicolumn{1}{c}{MJD} &
\multicolumn{2}{c}{$\left<B_{\mathrm z}\right>_{\rm all}$} &
\multicolumn{2}{c}{$\left<B_{\mathrm z}\right>_{\rm hydr}$} &
\multicolumn{1}{c}{Comment} \\
\multicolumn{1}{c}{} &
\multicolumn{1}{c}{} &
\multicolumn{2}{c}{[G]} &
\multicolumn{2}{c}{[G]} &
\multicolumn{1}{c}{} \\
\hline
3379 & 54109.048 & $-$53 &45 &$-$46  & 62& \\
     & 54112.025 & 30 & 28 & 85 & 55 & \\
     & 54344.233 & {\bf 117} &{\bf 34} & 67 & 38 & CD \\
     & 54345.189 & {\bf 155} & {\bf 42} & 58 & 52 & CD \\
11462 & 54344.248 & {\bf 161} & {\bf 46} & 132 & 48& ND \\
23958 & 54344.397 & 39 & 41 & 9 & 48 &  \\
24587 & 54086.175 & {\bf $-$353} & {\bf 82} & {\bf $-$329} & {\bf 91} &ND  \\
      & 54343.301 & 67 & 60 & 83 & 65 &  \\
25558 & 54086.242 & $-$71 & 43 & $-$105 & 48  &  \\
      & 54345.264 & {\bf 105} & {\bf 34} & 103 & 41& ND \\
26326 & 54086.263 & $-$30 & 33 & $-$42& 60&  \\
26739 & 54344.283 & 32 & 34 & 39 & 36&   \\
27742 & 54345.355 & 63 & 42 & 5 & 49&  \\
28114 & 54106.091 & {\bf 107} & {\bf 33} & 100 & 44& ND  \\
28475 & 54107.129 & 32 & 38 & 26 & 43 &  \\
      & 54345.312 & 94 & 36 & {\bf 160} & {\bf 48} & ND  \\
29376 & 54345.279 &100 &42  & 77 & 48&  \\
33331 & 54344.423 &$-$80 &38 &$-$21 & 42&  \\
34798 & 54100.150 & $-$99 & 45 & $-$106  & 50 \\
37151 & 54107.154 & $-$84 & 40 & $-$68 & 43&  \\
39844 & 54344.411 & $-$64 & 26  &$-$56  & 28 &  \\
40494 & 54343.426 &{\bf 94} & {\bf 28} & 39 & 40& ND \\
45284 & 54107.255 & $-$55 & 50 & $-$64 & 53&   \\
53921 & 54061.304 & $-$13 & 112 & $-$90 & 117&  \\
55718 & 54343.401 &$-$2  &41  &$-$21  &46 &   \\
69144 & 54061.342 &$-$37  &52  & $-$64 & 60 &   \\
74195 & 54108.330 &$-$102& 38 &$-$96  &43& \\
74560 & 54108.348 &{\bf $-$198}  & {\bf 55} &{\bf $-$191}  &{\bf 58}& CD \\
85953 & 54156.096 & 78 &27  & {\bf 97} & {\bf 29}& CD		 \\
128585 & 54344.976 &112 &51 &100 & 57 &  \\
140873 & 54179.299 &$-$144  & 51  &$-$173  &62&	 \\ 
       & 54344.011 & {\bf 99} & {\bf 31} & 51 & 31& CD 	\\ 
152511 & 54344.116 & {\bf 649} &{\bf 43}  & {\bf 728} & {\bf 50} & ND	\\
       & 54608.158 & {\bf 141}  &{\bf 26}  &{\bf 149}  & {\bf 31}& ND	  \\
       & 54609.433 & {\bf 440}  &{\bf 39} &{\bf 448} & {\bf 43} & ND		\\ 
       & 54610.223 &{\bf 158}  & {\bf 28}  &{\bf 173} & {\bf 31} &ND	\\
152635 & 54344.041 &{\bf $-$149}  &{\bf 36}  &{\bf $-$167}  & {\bf 38} & ND \\
163254 & 54344.068 & {\bf 155} &{\bf 49}  &81  &60 & ND		\\
169467 & 54345.164 & {\bf $-$182}& {\bf 41} &{\bf $-$233} &{\bf 43}& ND  \\
169820 & 54345.123 &$-$80 & 43 &$-$35 &48&	 \\
179588 & 54343.134 &{\bf 158}  &{\bf 41}  &{\bf 184}  & {\bf 43}& ND	 \\
181558 & 54344.167 &{\bf $-$104}  & {\bf 32} & $-$103  &  36 & CD   \\
183133 & 54344.179 & {\bf 152} & {\bf 38} & {\bf 200} & {\bf 43} & ND  	 \\
191295 & 54343.181 & 57 & 38 &23  &41 &  \\
       & 54345.218 & 102 &36  &39  &40 &		 \\
205879 & 54343.226 & {\bf 150} &{\bf 40}  & {\bf 156} & {\bf 43} & ND\\
206540 & 54344.220 &2  & 27 &$-$5  & 29 &	 \\
215573 & 54042.020 &137  & 52 & 129 & 57& 	 \\
       & 54343.244 &$-$7  &26  &$-$28  &29 &	 \\
       & 54345.232 &66  &33  &97  &38&	  \\ 
\hline
\end{tabular}
\end{center}
\end{table}

\addtocounter{table}{-1}
\addtocounter{subtable}{1}

\begin{table}
\caption{
Same as in Table~\ref{table2a} for the observed normal B stars and the N-rich star HD\,52089.
}
\label{table2c}
\begin{center}
%\begin{tabular}{rcrp{2mm}rcr}
\begin{tabular}{rcr@{$\pm$}rr@{$\pm$}rc}
\hline
\multicolumn{1}{c}{\raisebox{2mm}{\rule{0mm}{2mm}}HD} &
\multicolumn{1}{c}{MJD} &
\multicolumn{2}{c}{$\left<B_{\mathrm z}\right>_{\rm all}$} &
\multicolumn{2}{c}{$\left<B_{\mathrm z}\right>_{\rm hydr}$} &
\multicolumn{1}{c}{Comment} \\
\multicolumn{1}{c}{} &
\multicolumn{1}{c}{} &
\multicolumn{2}{c}{[G]} &
\multicolumn{2}{c}{[G]} &
\multicolumn{1}{c}{} \\
\hline
24626 & 54086.134 & 10 & 53 & 47& 59&   \\
52089 & 54046.339 & {\bf $-$200} & {\bf 48} & {\bf $-$271} & {\bf 51}&ND \\
      & 54343.389 &{\bf $-$129} &{\bf 34}  &{\bf $-$156}  &{\bf 18} &ND \\
142378 & 54344.025 &129  & 52  &131  & 62&	  \\
153716 & 54344.057 &{\bf 124}  & {\bf 41} &113  &43& ND  	\\
164245 & 54345.138 &116 & 40&98 &46 & \\
166197 & 54345.153 &$-$69 & 46 &$-$87  & 58 &	 \\ 
169033 & 54344.143 & 77 &45  & 99 & 48 &  	 \\
\hline
\end{tabular}
\end{center}
\end{table}
%THIERRY
%I would rename Tables 4-5 in Tables 2a-b.
%When there are more than one entry for a given star, I would give the HD number for the first entry only (to ease readability).
%You quote field strengths based on the analysis of all lines, but also only based on the Balmer lines. I guess we should make clear why we do so and what we regard as the most reliable values. There are sometimes large differences between the two (e.g. Fig.3). You preferably use the values derived from the Balmer lines in the figures (e.g. Fig.1), so my understanding is that <B_z>_hydr is more reliable. If so, there are several stars (e.g. HD 180642, HD 25558) with only detections in <B_z>_all, not <B_z>_hydr. Are we sure we can claim a detection for them? I see this as an important point.
%%%%%%%%%%%%

\section{Results}
\label{sect:results}

The results of our determinations of the mean longitudinal
magnetic field $\left<B_z\right>$ for all studied SPB stars, $\beta$\,Cephei stars, candidate SPB and 
$\beta$\,Cephei stars, and normal B-type stars are 
presented in Tables~\ref{table2a} to \ref{table2c}.
In the first two columns we give the HD number
and the modified Julian date of the middle of the exposures.
The measured mean longitudinal magnetic field $\left<B_{\mathrm z}\right>_{\rm all}$ using all absorption 
lines in the Stokes V spectra and $\left<B_{\mathrm z}\right>_{\rm hydr}$ measured on the hydrogen Balmer 
lines are presented in Cols.~3 and 4, respectively. 
As an important step, before the assessment of the
longitudinal magnetic field, we removed
all spectral features not belonging to the stellar photospheres of  the studied stars:
telluric and interstellar features,
CCD defects, also emission lines and lines with strong P\,Cygni profiles. 
%THIERRY
%The captions of the tables are often repeated in the text. Removing them would avoid repetition and save space.
%%%%%%%%%%%%
%In many cases more than one measurement has been carried out per star. 
For the early-type  $\beta$\,Cephei pulsators, a longitudinal magnetic field at a level larger than 3$\sigma$ has been
diagnosed in four $\beta$\,Cephei stars, $\delta$\,Cet, $\xi^1$\,CMa, HD\,50707, and HD\,180642, and two stars 
suspected to be 
$\beta$\,Cephei stars, HD\,74575 and HD\,136504.
For the SPB stars, a weak magnetic field has been diagnosed in
ten SPB stars, HD\,3379,  HD\,24587, HD\,25558, HD\,28114, HD\,28475, HD\,74560, HD\,85953, HD\,140873,
HD\,179588, and HD\,181558, and eight stars suspected to be SPB, HD\,11462, HD\,40494, HD\,152511,
HD\,152635, HD\,163254, HD\,169467, HD\,183133, and HD\,205879. 
Out of the ten detections among SPB stars, five are confirmed detections in the stars HD\,3379, HD\,74560, HD\,85953,
HD\,140873, and HD\,181558. These stars have already been studied between 2003 and 2005 and the detected magnetic fields 
were announced in our previous publication (Hubrig et al.\ \cite{Hubrig2006}).
We could not confirm the presence of magnetic fields at the
3$\sigma$ level for the previously studied five SPB stars, HD\, 45284, HD\,53921, HD\,74195, HD\,169820, 
and HD\,215573.
These non-detections are likely caused by the strong dependence of the longitudinal magnetic field on rotational
aspect. It is generally known that the usefulness of longitudinal magnetic fields in characterizing actual 
magnetic field strength distributions depends on the sampling of various rotation phases,
and hence various aspects of the magnetic field. 
Among seven normal B-type stars not known as pulsating stars, a weak magnetic field was detected in 
the nitrogen-rich early B-type star HD\,52089 and in the B5 IV star HD\,153716.
%THIERRY
%You refer to the the same star in the text either using its usual name or its HD number (e.g. xi^1 CMa or HD 46328). It's a bit confusing.
%%%%%%%%%%%%

\begin{figure}
\centering
\includegraphics[width=0.35\textwidth]{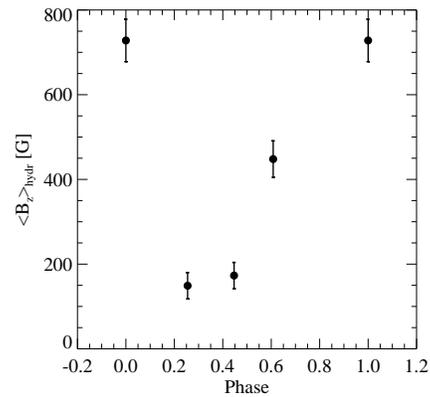}
\caption{
Longitudinal magnetic field measurements with FORS\,1 using hydrogen lines in the 
candidate SPB star HD\,152511 over the period of 0.94\,d.
}  
\label{fig:152511phase}  
\end{figure}

Among the SPB stars, the detected magnetic fields are mainly of the order of 100--200\,G. Only the measurements 
in HD\,24587 
revealed the presence of a comparatively large magnetic field of the order of 350\,G.
For stars observed more than once, the individual measurements show a variability of the magnetic field.
%Due to the strong dependence of the longitudinal magnetic field on rotational 
%aspect, its usefulness in characterizing actual magnetic field strength distributions
%depends on the sampling of the various rotation phases,
%and hence various aspects of the magnetic field. 
However, no exact rotation periods are known for the stars in our sample, 
and it is certainly not possible with just a few measurements to obtain a clue about the magnetic field 
geometry causing the observed variations.

A rather strong variable magnetic field has been detected in the candidate SPB star HD\,152511 with a 
maximal field strength of $\left<B_{\mathrm z}\right>_{\rm hydr}$\,=\,728$\pm$50\,G  measured on the hydrogen Balmer 
lines.
This star, however, is poorly studied, with only six references in the SIMBAD database.
Hipparcos photometric observations reveal a variation period of the order of 0.94\,d, which corresponds to the 
pulsation range of SPB stars. It is presently not clear whether this period is in fact a rotation period and the 
observed photometric variability 
%should 
%THIERRY
could
%%%%%%%%%%%%
 be attributed to an inhomogeneous distribution of chemical elements 
on the stellar surface. The magnetic field measurements show a positive longitudinal magnetic field over the 
period of 0.94\,d 
without any change of polarity (Fig.~\ref{fig:152511phase}).  This star obviously deserves future 
spectropolarimetric, spectroscopic and
photometric observations to establish the nature of its variability.

\begin{figure}
\centering
\includegraphics[width=0.45\textwidth]{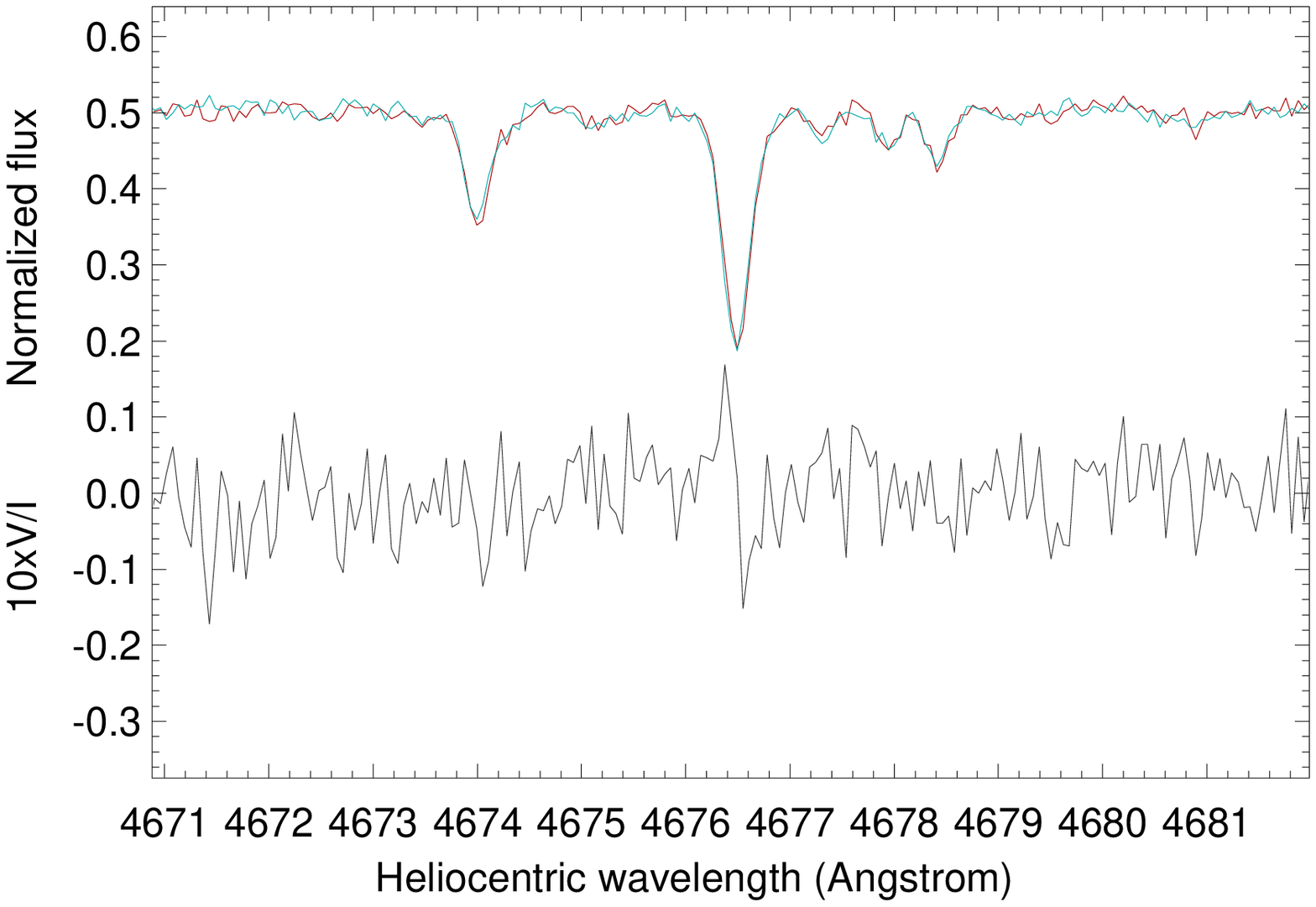}
\includegraphics[width=0.45\textwidth]{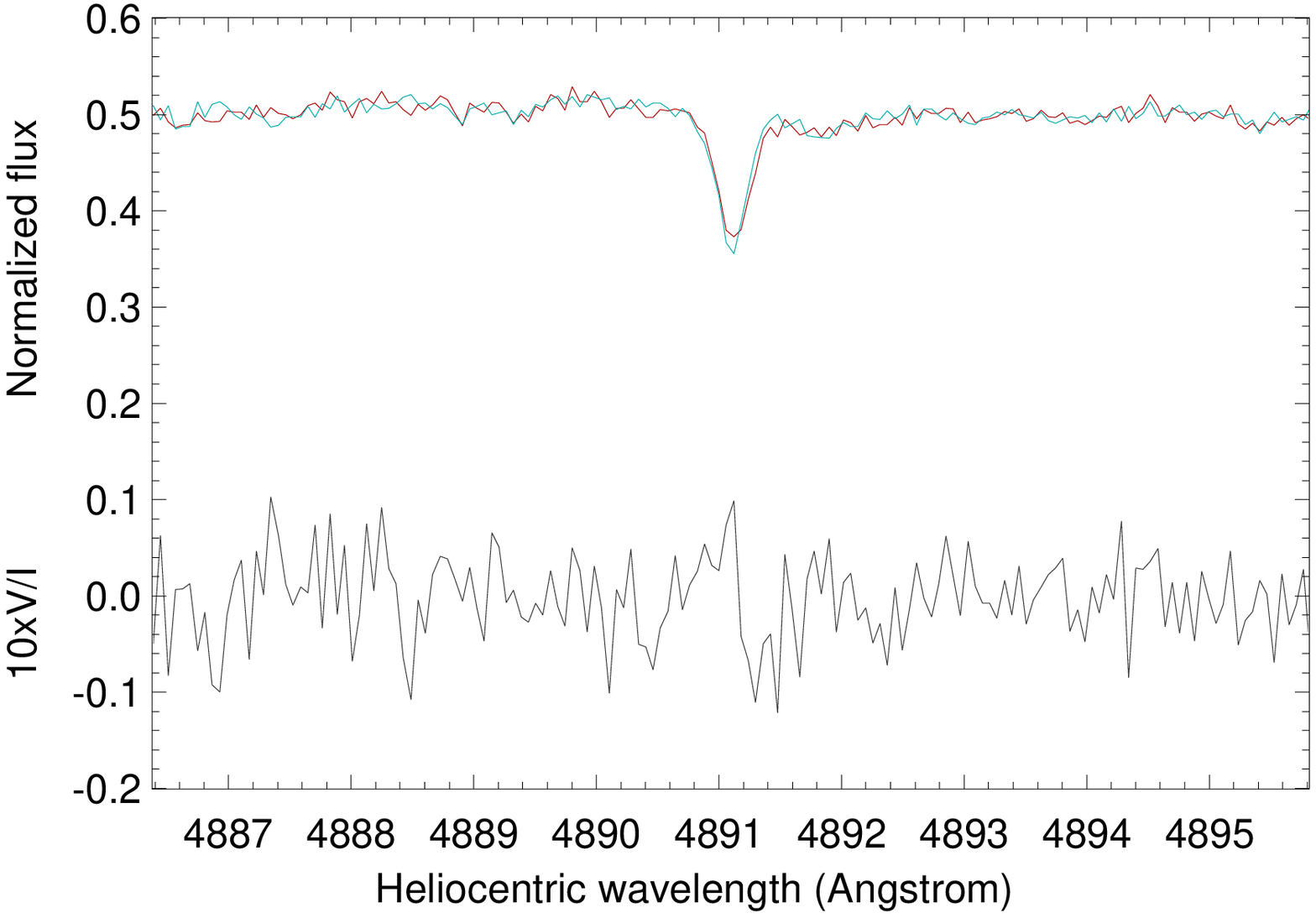}
\includegraphics[width=0.45\textwidth]{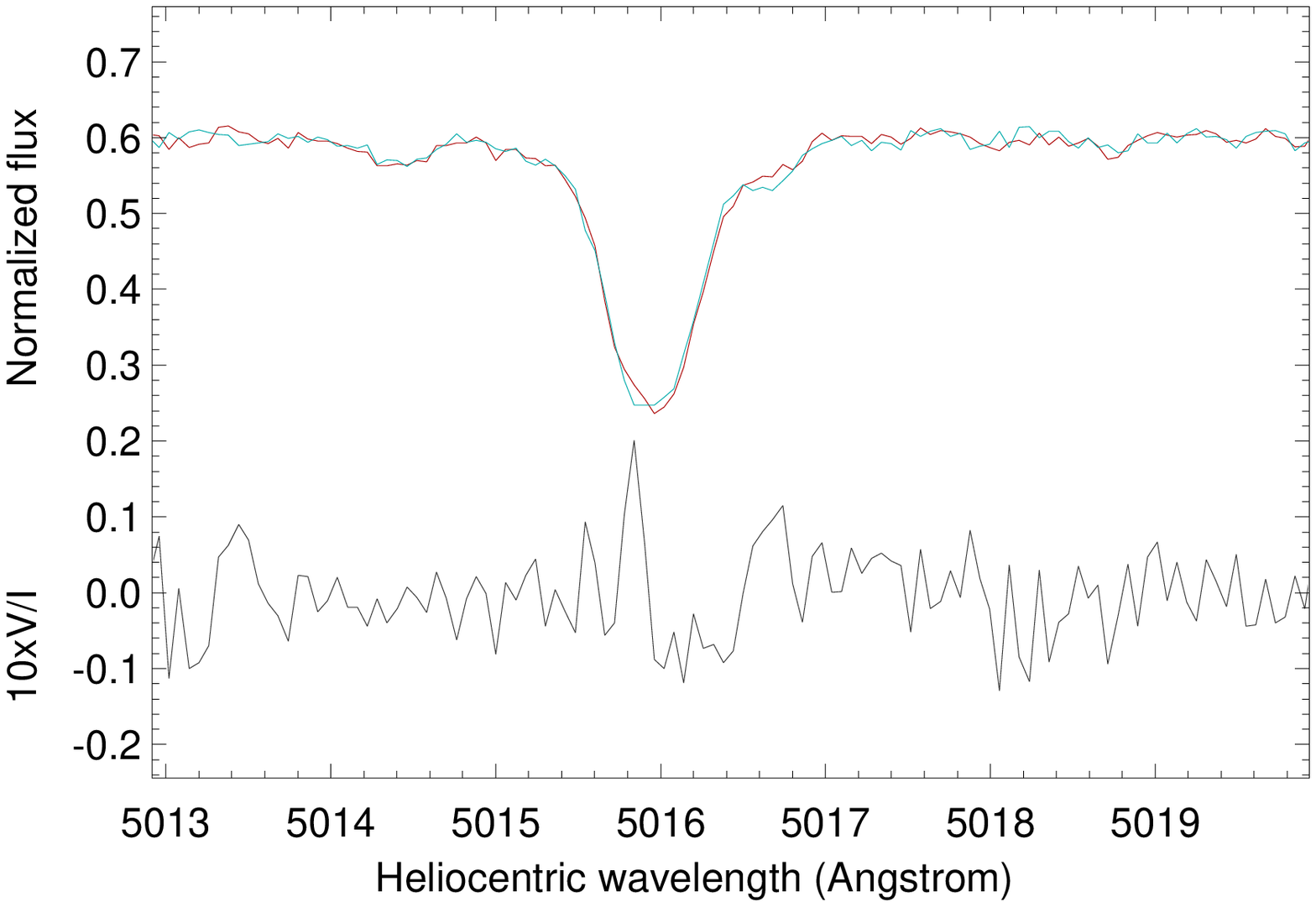}
\caption{High-resolution (R$\approx$30\,000) SOFIN polarimetric spectra of $\xi^1$\,CMa.
In each panel, the upper two (blue and red) are the polarized spectra,
while the lower spectrum is the deduced $V$/$I$ spectrum scaled by a factor 10.
Clear Zeeman features are detected 
at the positions of the unblended lines O~II 4676.2 and O~II 4890.9. The line He~I 5015.7 is marginally blended with the 
line N~II 5016.4.
}  
\label{fig:not}  
\end{figure}

\begin{figure}
\centering
\includegraphics[width=0.35\textwidth]{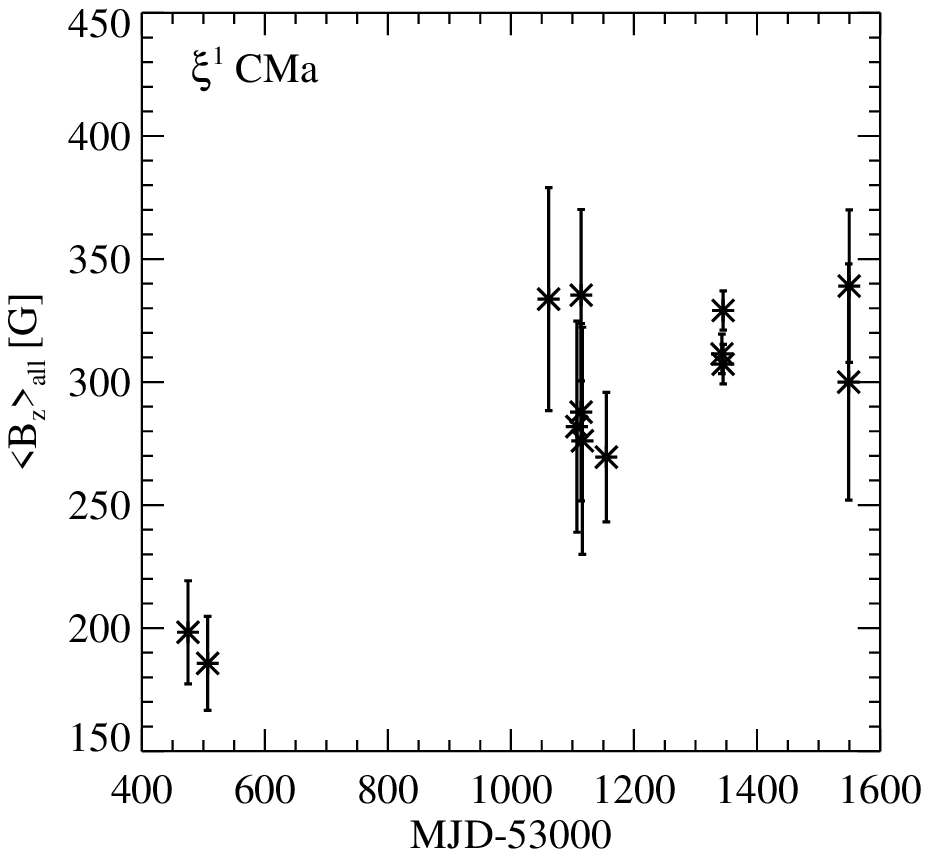}
\includegraphics[width=0.35\textwidth]{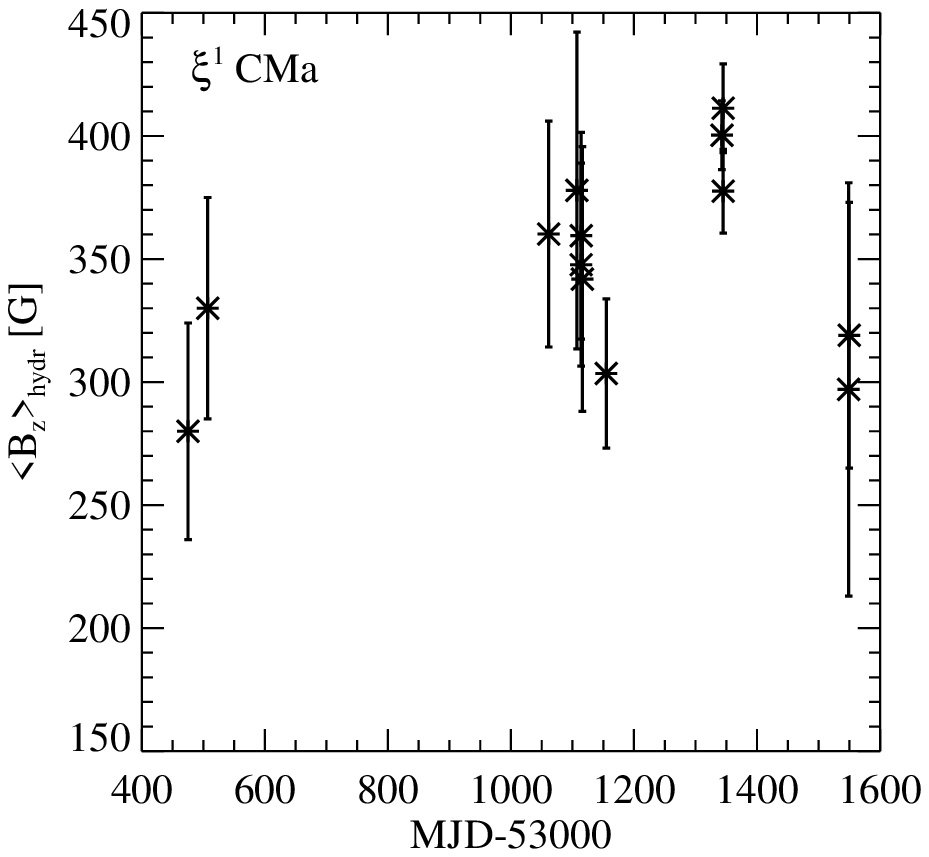}
\caption{Magnetic field measurements for $\xi^1$\,CMa in the last 4.4~years,
using all lines (upper panel) and only the hydrogen lines (lower panel).
}  
\label{fig1}  
\end{figure}
%THIERRY
%I would show a figure with two panels instead (top/bottom).
%%%%%%%%%%%%

Magnetic fields of the order of several ten to several hundred Gauss have been detected in four $\beta$\,Cephei 
stars and in two candidate $\beta$\,Cephei stars.
%Among the $\beta$\,Cephei pulsators, 
The presence of a relatively strong magnetic field  in the $\beta$\,Cephei 
star $\xi^1$\,CMa is confirmed 
in all eleven  measurements carried out since 2005. 
In Fig.~\ref{fig:not} we present our high resolution spectropolarimetric observations with the 
SOFIN echelle spectrograph at the Nordic Optical Telescope. 
In spite of rather strong noise, clear Zeeman features are detected 
at the positions of the unblended lines O~II 4676.2, O~II 4890.9, and He~I 5015.7.
The latter line is only marginally blended with N~II 5016.4. 
In Fig.~\ref{fig1} we present the acquired 
measurements of this star over the last 4.4~years with FORS\,1. No polarity change is 
detected in our measurements.
Measurements of the magnetic field 
during the same nights or within one day (MJDs 54114, 54345, 54348--54349) show small changes in the 
magnetic field strength of the order of a few tens of Gauss, indicating that the field is likely slightly variable on 
rather short time scales, a couple of days at most. 
%THIERRY
%Are you sure these changes are real? To me, they are identical within the errors bars.
%%%%%%%%%%%%
 It is not clear yet whether this variability is caused by 
stellar rotation or by stellar pulsations. 
On the other hand, from the measured $v$\,sin\,$i$ value and the radius estimation presented in Table~\ref{table1a}
we derive a rotation period in the range from 12 to 33\,days. 
As we reported in our previous study (Hubrig et al.\ \cite{Hubrig2006}),
the radial velocity variability of this star was first discovered by Frost (\cite{Frost1907}).
This star pulsates in a radial mode monoperiodically (Saesen et al.\ \cite{Saesen2006}) with a period of 0.209574 days
(e.g., Heynderickx et al.\ \cite{Heynderickx1994}). 
Saesen et al.\ (\cite{Saesen2006}) found a peak-to-peak radial-velocity amplitude of some 33 km~s$^{-1}$,
which is among the largest values observed for a $\beta$\,Cephei star.
The amplitude of magnetic field variations is rather low: The obtained values of the longitudinal magnetic 
fields are in the range from 308 to 380\,G for the measurements 
using all spectral lines and from 276 to 400\,G for the measurements using hydrogen lines.
Our search for a variation period of the magnetic field in $\xi^1$\,CMa using a
Fourier analysis could not reveal any significant frequency.

It is intriguing that the distribution of pulsation periods of early B-type 
pulsating stars is similar to the distribution of the rotation periods of a number of chemically peculiar magnetic He-strong
stars which occupy approximately the same parameter space in the H-R diagram as 
$\beta$\,Cephei and SPB stars (e.g., Briquet et al.\ \cite{Briquet2007}).
The photometric, spectroscopic and magnetic field variations of He-strong stars are usually 
interpreted in terms of the oblique rotator model. 
Several such stars, e.g., HD\,36485, HD\,58260, HD\,60344, HD\,96446, and
HD\,133518, with rather 
short periods, show no polarity change and rather low 
variability amplitude of the longitudinal magnetic field similar to the magnetic field behaviour of  $\xi^1$\,CMa.
For the He-strong star
HD\,96446 Mathys (\cite{Mathys1994})  obtained an unrealistic small radius from the study of the observed magnetic 
structure, suggesting that the observed variability could be caused by pulsations. 
Interestingly, all four 3$\sigma$ magnetic field detections in the $\beta$\,Cephei star $\delta$~Cet 
(Aerts et al.\ \cite{Aerts2006}) indicate a negative magnetic field without any
change of polarity. 
%Two other measurements show positive polarity but due to the low significance of less than 2$\sigma$ they
%cannot be considered as real detections. 
%The magnetic field $\delta$~Cet is very faint, with the average value around 40-50~G. 
According to Aerts et al.\ (\cite{Aerts2006}) $\delta$~Cet is most likely observed nearly pole-on.
One more  $\beta$\,Cephei star in our sample, HD\,50707 does not show any change of polarity either, though 
only two measurements have been carried out so far for this star.
% we do not wish to forejudge about the resemblance between these objects and $\xi^1$\,CMa.

Recently, Morel et al.\ (\cite{Morel2006,Morel2008}) performed 
%an abundance analysis in 
%THIERRY
an NLTE abundance analysis of
%%%%%%%%%%%%
 a sample of slowly rotating early-type B dwarfs with detected weak magnetic fields.
The studied sample 
included, among other stars, also a number of SPB and  $\beta$\,Cephei stars for which we carried out 
a magnetic field survey in recent years.
%Noteworthy, a NLTE abundance analysis 
%THIERRY
This analysis
%%%%%%%%%%%%
strongly supports the existence of a population of nitrogen-rich and boron-depleted slowly rotating B stars
%Morel et al.\ \cite{Morel2006}  discovered that 
%THIERRY
 and indicates that 
%%%%%%%%%%%%
 the $\beta$\,Cephei stars $\delta$~Cet, $\xi^1$\,CMa,
HD\,50707, and the early B-type star  HD\,52089 are all nitrogen enriched. For all these stars 
we have magnetic field detections.
In addition, two other $\beta$\,Cephei stars, 
%2052\,Oph
%THIERRY
 V2052\,Oph
%%%%%%%%%%%% 
 and $\beta$\,Cep, of their sample with detected 
longitudinal magnetic fields of the order of $\sim$100\,G (Neiner et al.\ \cite{Neiner2003b}; Henrichs et al.\ \cite{Henrichs2000}) were 
found to show nitrogen enhancement. 
For the sixth  $\beta$\,Cephei star with a detected magnetic field, HD\,180642,
Morel \& Aerts (\cite{MorelAerts2007}) found a mild nitrogen overabundance,
though boron data are not available.
Also for the candidate $\beta$~Cephei star HD\,74575 with diagnosed magnetic field,  Przybilla et al.\ (\cite{Przybilla2008})
recently detected nitrogen overabundance,
while Profitt \& Quigley (\cite{ProffittQuigley2001}) found this star to be boron-depleted.
%For these both stars, HD\,74575 and HD\,180642, we achieved a magnetic field detection at the 3$\sigma$ level.
Unfortunately, nothing is known about the abundances of these chemical elements
for the remaining candidate $\beta$~Cephei star HD\,136504 with a detected magnetic field.
In summary, all confirmed and candidate $\beta$~Cephei stars,
with a magnetic field detection and with available N and B abundance values, present a nitrogen enrichment 
accompanied by a boron depletion. 

\addtocounter{table}{0}
\addtocounter{subtable}{-3}

\begin{table}
\caption{Logarithmic ratio of the nitrogen and carbon abundances ([N/C]) for the $\beta$\,Cephei stars, SPBs and non-pulsating B stars. 
An asterisk denotes a candidate $\beta$\,Cephei or SPB star, and stars in boldface have a magnetic field detection.
The solar [N/C] abundance ratio is about $-$0.6\,dex.}
\label{tab:nc}
\begin{center}
\begin{tabular}{rccc}
\hline
\hline
\multicolumn{1}{c}{HD} &
\multicolumn{1}{c}{Other} &
\multicolumn{1}{c}{[N/C]} &
\multicolumn{1}{c}{Ref.} \\ 
\multicolumn{1}{c}{} &
\multicolumn{1}{c}{Identifier} &
\multicolumn{1}{c}{[dex]} &
\multicolumn{1}{c}{} \\
\hline
{\bf        16582}  & {\bf $\delta$~Cet}  & {\bf --0.04$\pm$0.14} & {\bf 1}\\
{\rm        29248}  & {\rm $\nu$~Eri}     & {\rm --0.37$\pm$0.15} & {\rm 1}\\
{\rm        44743}  & {\rm $\beta$~CMa}   & {\rm --0.57$\pm$0.18} & {\rm 1}\\
{\bf        46328}  & {\bf $\xi^1$~CM}a   & {\bf --0.18$\pm$0.21} & {\bf 1}\\
{\bf        50707}  & {\bf 15~CMa}        & {\bf --0.15$\pm$0.19} & {\bf 1}\\
{\rm        61068}  & {\rm PT~Pup}        & {\rm --0.15$\pm$0.12} & {\rm 2}\\
{\bf$\ast$  74575}  & {\bf $\alpha$~Pyx}  & {\bf --0.27$\pm$0.13} & {\bf 3}\\
{\rm       111123}  & {\rm $\beta$~Cru}   & {\rm --0.43$\pm$0.20} & {\rm 1}\\
{\rm       129557}  & {\rm BU~Cir }       & {\rm  +0.12$\pm$0.13} & {\rm 2}\\
{\rm       129929}  & {\rm V836~Cen}      & {\rm --0.61$\pm$0.17} & {\rm 1}\\
{\bf       180642}  & {\bf V1449~Aql}     & {\bf --0.21$\pm$0.22} & {\bf 4}\\
\hline
{\bf        85953}  & {\bf V335~Vel}      & {\bf --0.50$\pm$0.24} & {\bf 5}\\    
{\bf$\ast$ 169467}  & {\bf $\alpha$~Tel}  & {\bf --0.11$\pm$0.20} & {\bf 6}\\  
{\bf$\ast$ 215573}  & {\bf $\xi$~Oct}     & {\bf --0.48}          & {\bf 7}\\    

\hline
{\bf        52089}  & {\bf 21~CMa}        & {\bf --0.16$\pm$0.19} & {\bf 1}\\  
\hline
\end{tabular}
\begin{flushleft}
Key to references:
[1] Morel et al.\  (\cite{Morel2008}),
[2] Kilian (\cite{Kilian1992}),
[3] Przybilla et al.\  (\cite{Przybilla2008}), 
[4] Morel \& Aerts (\cite{MorelAerts2007}),
[5] Briquet \& Morel (\cite{BriquetMorel2007}),
[6] Zboril \& North (\cite{ZborilNorth1999}), and
[7] Pintado \& Adelman (\cite{PintadoAdelman2003}).
\end{flushleft}
\end{center}
\end{table}

%At present, only for three SPBs and candidate SPBs, all of early B spectral type, the nitrogen abundance is known.
%Two SPB stars with detected magnetic fields, $\zeta$~Cas and HD\,85953 have recently been studied by 
%citet\cite{Morel2008}. While $\zeta$~Cas was found nitrogen deficient, 
% the analysis for HD\,85953 revealed a rather normal nitrogen composition. For the candidate SPB star HD\,169467, 
%with the measured longitudinal magnetic field $\left<B_{\mathrm z}\right>_{\rm hydr}$\,=\,$-$233$\pm$43\,G,
%Zboril \& North (\cite{ZborilNorth1999}) determined ratios of the C, N and O abundances ([N/C] and [N/O]),
%which are typical of those found by Morel et al.\ \cite{Morel2008} in their sample of magnetic slowly-rotating B-type dwarfs,
%indicating that this star is N-rich too.
%Boesgaard \& Heacox (\cite{BoesgaardHeacox1978}) used Copernicus observations to determine the boron abundance in a sample of 
%B-type stars and have
%found the HD\,169467 boron depleted, assigning the same boron abundance to this star as to the magnetic SPB star $\zeta$~Cas.
%The more recent and more reliable analysis of $\zeta$~Cas by Profitt \& Quigley (\cite{ProffittQuigley2001}) confirmed the very low boron abundance
%in the magnetic SPB star $\zeta$~Cas, and, consequently, HD\,169467 should be boron deficient too.
%THIERRY
At present, only for three confirmed and candidate SPBs, all of early B spectral type, is the nitrogen abundance known.
Two SPB stars with detected magnetic fields, $\zeta$~Cas and HD\,85953 have recently been studied by Briquet \& Morel (\cite{BriquetMorel2007}). While $\zeta$~Cas was found to be nitrogen rich, 
 the analysis of HD\,85953 revealed a normal chemical composition. For the candidate SPB star HD\,169467, 
with a measured longitudinal magnetic field $\left<B_{\mathrm z}\right>_{\rm hydr}$\,=\,$-$233$\pm$43\,G,
Zboril \& North (\cite{ZborilNorth1999}) determined LTE abundance ratios of C, N and O ([N/C] and [N/O])
that are typical of those found by Morel et al.\ (\cite{Morel2008}) in their sample of magnetic slowly-rotating B-type dwarfs,
indicating that this star is N-rich too.
Boesgaard \& Heacox (\cite{BoesgaardHeacox1978}) used Copernicus observations to determine the boron abundance in a sample of 
B-type stars and found HD\,169467 to be boron depleted, assigning the same boron abundance to this star as to the magnetic SPB star $\zeta$~Cas.
The more recent and more reliable analysis of $\zeta$~Cas by Profitt \& Quigley (\cite{ProffittQuigley2001}) confirmed the very low boron abundance
in this magnetic SPB star, and, consequently, that HD\,169467 could be boron poor at similar levels.
An overview of the available abundance analyses of nitrogen in pulsating B-type stars 
is given in Table~\ref{tab:nc}, where we present nitrogen over carbon abundance ratios ([N/C]).
Almost all stars with a nitrogen overabundance have detected magnetic fields.
Unfortunately, we have only one magnetic field measurement for both HD\,61068 and, even more 
regrettably, HD\,129557. The latter is one of the most N-rich B stars known in the solar neighbourhood.

Clearly, the presently available observational data suggest a higher incidence of chemical 
peculiarities in stars with detected magnetic fields. We note that $\delta$~Cet, HD\,50707, and 
HD\,52089 have been included in this survey primarily because they were nitrogen rich. These results 
open a new perspective for the selection of
the most promising targets for magnetic field surveys of massive stars using chemical anomalies
as selection criteria.
%THIERRY
%Perhaps we can say here that delta Cet, HD 50707 and HD 52089 have been included in this survey primarily because they were N rich, so an N excess seems to be a pretty good criterion to select magnetic stars.
%%%%%%%%%%%%

%Since the presence of a magnetic field in these stars might play
%an important role to explain these physical 
%characteristics we studied the distribution of frequencies in stars with detected 
%magnetic fields. 

\section{Discussion}

\begin{table*}
\caption{The frequencies and the corresponding photometric pulsating amplitudes of all pulsating stars for 
which magnetic fields have been detected with FORS\,1.
An asterisk denotes candidate $\beta$\,Cep and SPB stars.
}
\label{table_freq}
\begin{center}
\begin{tabular}{rcccccccc}
\hline
\hline
\multicolumn{1}{c}{HD} &
\multicolumn{1}{c}{$f_1$} &
\multicolumn{1}{c}{$a_1$} &
\multicolumn{1}{c}{$f_2$} &
\multicolumn{1}{c}{$a_2$} &
\multicolumn{1}{c}{$f_3$} &
\multicolumn{1}{c}{$a_3$} &
\multicolumn{1}{c}{$f_4$} &
\multicolumn{1}{c}{$a_4$} \\
&
\multicolumn{1}{c}{[c/d]} &
\multicolumn{1}{c}{[mmag]} &
\multicolumn{1}{c}{[c/d]} &
\multicolumn{1}{c}{[mmag]} &
\multicolumn{1}{c}{[c/d]} &
\multicolumn{1}{c}{[mmag]} &
\multicolumn{1}{c}{[c/d]} &
\multicolumn{1}{c}{[mmag]} \\
&
\multicolumn{1}{c}{$f_5$} &
\multicolumn{1}{c}{$a_5$} &
\multicolumn{1}{c}{$f_6$} &
\multicolumn{1}{c}{$a_6$} &
\multicolumn{1}{c}{$f_7$} &
\multicolumn{1}{c}{$a_7$} &
\multicolumn{1}{c}{$f_8$} &
\multicolumn{1}{c}{$a_8$} \\
&
\multicolumn{1}{c}{[c/d]} &
\multicolumn{1}{c}{[mmag]} &
\multicolumn{1}{c}{[c/d]} &
\multicolumn{1}{c}{[mmag]} &
\multicolumn{1}{c}{[c/d]} &
\multicolumn{1}{c}{[mmag]} &
\multicolumn{1}{c}{[c/d]} &
\multicolumn{1}{c}{[mmag]} \\
\hline
\multicolumn{9}{c}{$\beta$\,Cep stars} \\
\hline
       16582  & 6.20589(M)     &11.6 & 3.737(M)  &  0.53& 3.672(M) &   0.39& 0.318(M)  &  0.68 \\
       46328  & 4.77153(GHSp)  &32.4 \\
$\ast$ 50707  & 5.4187(GHStSp)& 6.2 & 5.1831(St) &  5.4 & 5.5212(St)&   2.0 & 5.3085(St) &  1.6 \\
%$\ast$ 74575  \\
$\ast$ 136504 &  10.36(Sp)      \\
       180642 &  5.4869(G) & 82.8 & 0.3082(G) & 5.2 & 7.3667(G) & 4.6  \\
\hline
\multicolumn{9}{c}{SPB stars} \\
\hline
       3379  &  1.82023(G)    &  5.0 & 1.59418(G)  &  4.2 \\
$\ast$ 11462 &  1.33482(H)    & 18.0 & 1.44486(H)  & 12.0 \\
       24587 &  1.1569(GHSp) & 13.1 \\                                                                                          
       25558 &  0.65265(GH)   & 26.3 & 1.93235(GH) &  6.2 & 1.17913(G) &   5.1 \\
       28114 &  0.48790(G)    & 26.2 & 0.48666(G)  & 12.9 & 0.79104(H) \\
       28475 &  0.68369(GH)   & 14.9 & 0.40893(G)  &  8.8 \\
%$\ast$ 40494 \\  
       45284 &  1.23852(G)    & 13.3 & 1.12753(G)   &19.5 & 1.50605(G)   & 9.1 & 1.32975(H)   & 9.7 \\
       53921 &  0.6054(GHSp) & 12.8 \\
       74195 &  0.35745(GHSp) & 23.4 & 0.35033(GHSp)&18.8 & 0.34630(GHSp)& 6.0 & 0.39864(GHSp)& 4.5 \\
       74560 &  0.64472(GHSp) & 25.0 & 0.39578(G Sp)& 7.0 & 0.44763(G)   & 6.0 & 0.82281(GH)  & 5.4 \\
             & 0.63567(GH) &  5.7 \\
       85953 &  0.2663(GHSp) & 14.7 & 0.2189(GH)  & 5.1 & 0.2353(Sp) \\
       140873 &  1.1515(GHSp)&  28.4 \\
       %143309 &  0.59830(G)   &  44.0 & 0.60213(G)  & 36.9 & 0.71414(G) &  21.3 & 1.20066(G) &  11.5 \\
              %& 0.59133(G) &  11.4 \\
$\ast$ 152511 &  1.0614(H)   &  14.3 \\
$\ast$ 152635 &  0.5211(H)   &  20.4 \\
       160124 &  0.52014(G)   &  23.7 & 0.52096(G)  & 18.9 & 0.52055(G) &  23.5 & 0.52137(G) &  20.1 \\
              & 0.52197(G)    &  22.3 & 0.70259(G)  & 15.0 & 1.04220(G) &  10.7 & 0.31366(G) &  11.5 \\
       161783 &  0.7351(St)  &   6.5 & 0.5602(St) &  6.2 & 0.9254(St) &  3.0 & 0.3256(St) &  4.8 \\
$\ast$ 163254 &  1.20241(H)   &  24.9 \\
$\ast$ 169467 &  1.1003(H)   &   6.2 \\
       169820 &  2.12509(GH)  &  13.6 \\
       179588 &  0.85654(G)   &  29.7 & 2.04263(G) &  17.6 & 2.19989(G)  & 16.2 & 1.83359(G) &  13.5 \\
              &  0.81599(H)  & 18.4 \\
       181558 &  0.80780(GHSp)&  49.5 \\
$\ast$ 183133 &  0.98177(H)   &  16.9 \\
$\ast$ 205879 &  1.01729(H)   &  20.0 & 0.98251(H)   & 8.0 \\
       208057 &  0.89050(GH)  &  13.5 & 0.80213(GH)  & 9.4 & 2.47585(G) &   9.1 \\
$\ast$ 215573 &  0.5439(H)   &  11.5 & 0.5654(GHSp)&10.8 \\
%\hline
%\multicolumn{9}{c}{Other objects} \\
%\hline
       %52089  \\
       %153716 \\
\hline
\end{tabular}
\end{center}
\begin{flushleft}
(G) -- Geneva photometry \\
(H) -- Hipparcos photometry \\
(M) -- MOST Photometry \\
(Sp) -- Spectroscopic data \\
%MS: Shouldn't this be Str\"omgren ???
(St) -- Stromgren photometry \\
\end{flushleft}
\end{table*}

\begin{figure}
\centering
\includegraphics[height=0.48\textwidth,angle=270]{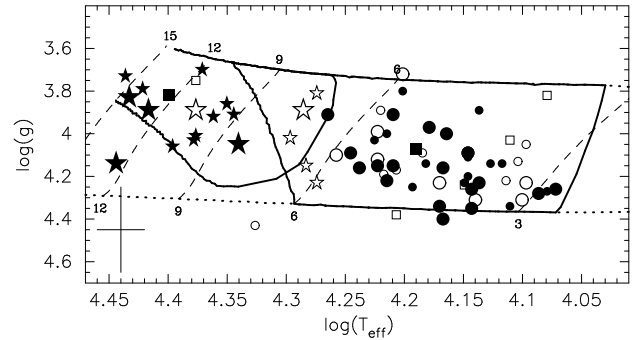}
\caption{
The position of the studied pulsating and non-pulsating stars in the H-R diagram. 
The sample includes all targets for which FORS\,1 spectropolarimetric observations 
are available (this study and Hubrig et al.\ \cite{Hubrig2006}).
The full lines represent the boundaries of 
the theoretical instability strips for modes with frequencies between 0.2 and 25\,d$^{-1}$ 
and $\ell \le 3$, computed for main-sequence models with 2\,$M_\odot$ $\le M \le$ 15\,$M_\odot$
(De Cat et al.\ \cite{DeCat2007a}).
The lower and upper dotted lines show the zero-age main sequence 
and terminal-age main sequence, respectively. The dashed lines denote evolution tracks for 
stars with $M$ = 15, 12, 9, 6, and 3\,$M_\odot$.
%Coloured symbols correspond to stars with detected magnetic fields, 
Filled circles correspond to confirmed SPB stars, open circles to 
candidate SPB stars, filled stars to confirmed $\beta$~Cephei stars, open stars to candidates $\beta$~Cephei stars, 
and squares to standard B stars. The stars with detected magnetic fields are presented 
by  symbols which are 1.5 times bigger than those for stars for which magnetic fields were not detected.
The cross in the bottom left corner gives the typical error estimate, 0.02 on log(\teff{}) and 0.2 on \logg{}.
The parameters of objects with M $>$ 12\,$M_\odot$ are less reliable because these values
were found by extrapolation out of the calibration tables.
}
\label{fig_peter}  
\end{figure}

\begin{figure}
\centering
\includegraphics[height=0.40\textwidth,angle=0]{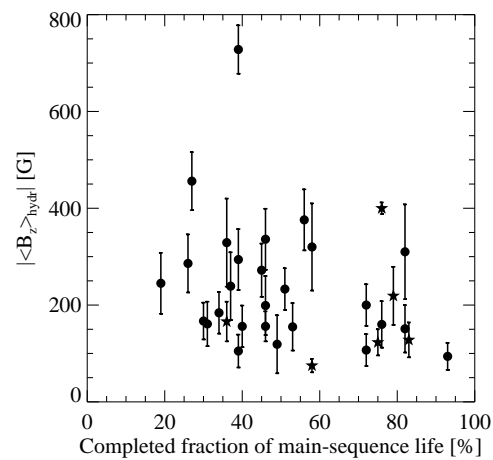}
\caption{
The strength of the longitudinal magnetic field measured with FORS\,1 using hydrogen lines
as a function of the completed fraction of the main-sequence lifetime.
Filled stars indicate $\beta$~Cephei stars and candidate $\beta$~Cephei stars,
while filled circles indicate SPBs and candidate SPBs.
}
\label{fig:f_Hhyd}
\end{figure}
%THIERRY
%At first I was confused as to why one magnetic SPB on the ZAMS (with log Teff~4.17) had f as high as 19%, but then I realised that actually f=19+/-11% so this is more or less OK (but still formally incompatible with 0 though...). The errors on f are quite large, perhaps we could put them on Fig.5?
%I guess this figure (and the others) would be clearer with filled/open symbols (and with colours if free...).
%%%%%%%%%%%%

\begin{figure}
\centering
\includegraphics[height=0.40\textwidth,angle=0]{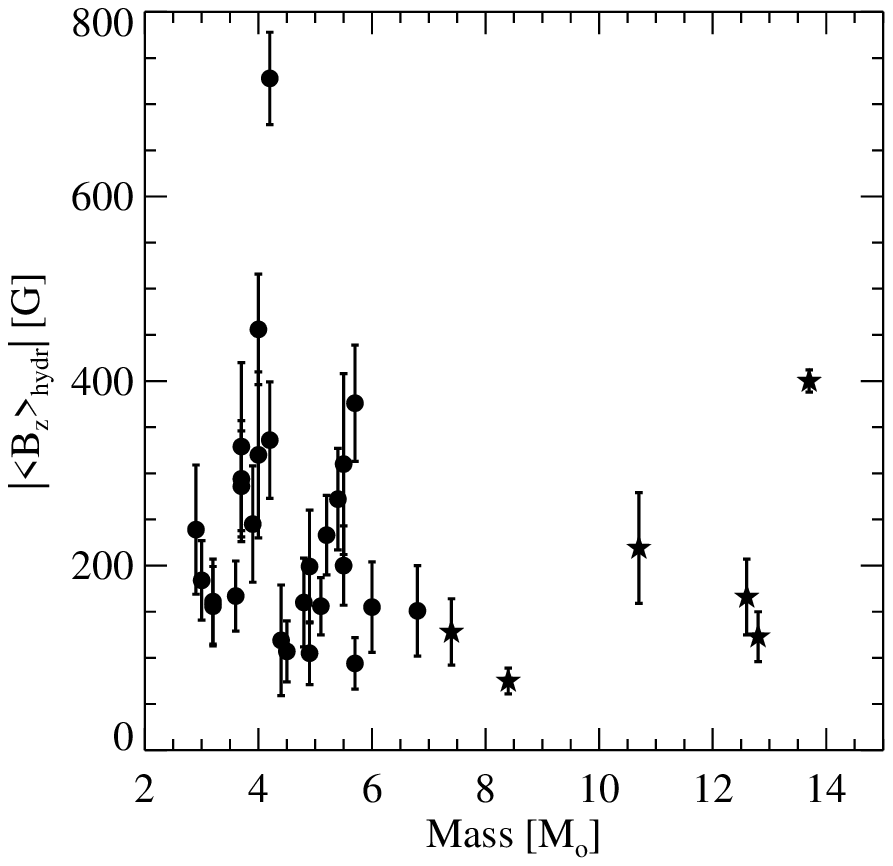}
\caption{
The strength of the longitudinal magnetic field measured with FORS\,1 using hydrogen lines
as a function.of stellar mass. The symbols are identical to those presented in Fig.~\ref{fig:f_Hhyd}.
}
\label{fig:M_Hhyd}
\end{figure}

\begin{figure}
\centering
\includegraphics[height=0.40\textwidth,angle=0]{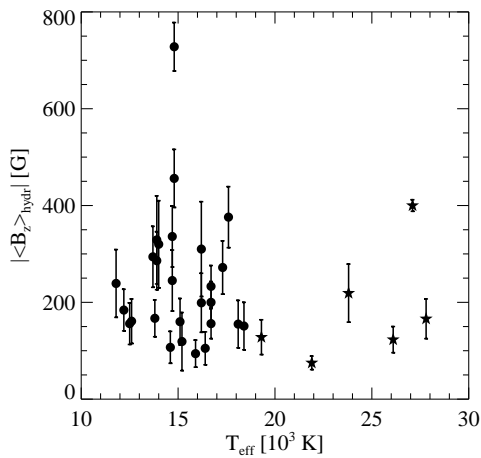}
\caption{
The strength of the longitudinal magnetic field measured with FORS\,1 using hydrogen lines
as a function of effective temperature. The symbols are identical to those presented in Fig.~\ref{fig:f_Hhyd}.
}
\label{fig:teff_Hhyd}
\end{figure}

\begin{figure}
\centering
\includegraphics[height=0.40\textwidth,angle=0]{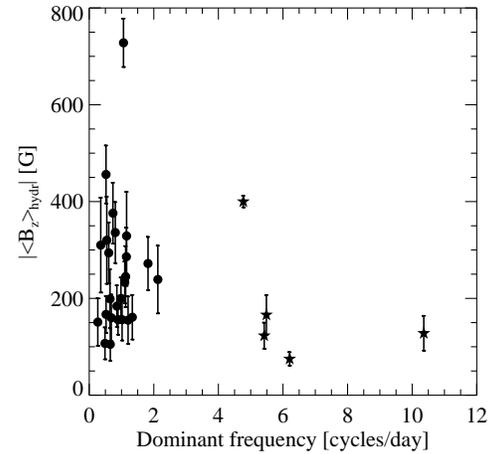}
\caption{
The strength of the longitudinal magnetic field measured with FORS\,1 using hydrogen lines
as a function of frequency presented in Table~\ref{table_freq}. 
For multiperiodic stars we selected the pulsation mode with the highest amplitude.
 The symbols are identical to those in Fig.~\ref{fig:f_Hhyd}.
}
\label{fig:freq_Hhyd}
\end{figure}

\begin{figure}
\centering
\includegraphics[height=0.40\textwidth,angle=0]{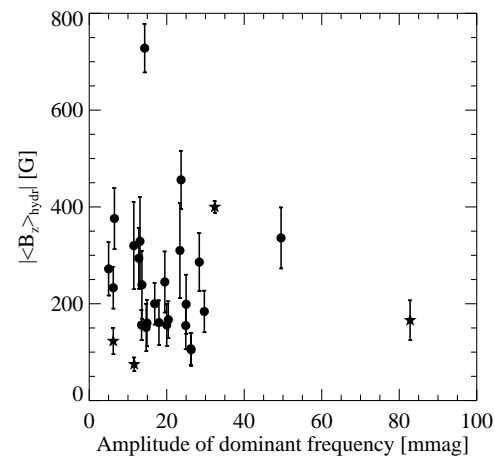}
\caption{
The strength of the longitudinal magnetic field measured with FORS\,1 using hydrogen lines
as a function of the pulsation amplitude of the dominant frequency.
The symbols are identical to those presented in Fig.~\ref{fig:f_Hhyd}.
}
\label{fig:amp_Hhyd}
\end{figure}

\begin{figure}
\centering
\includegraphics[height=0.40\textwidth,angle=0]{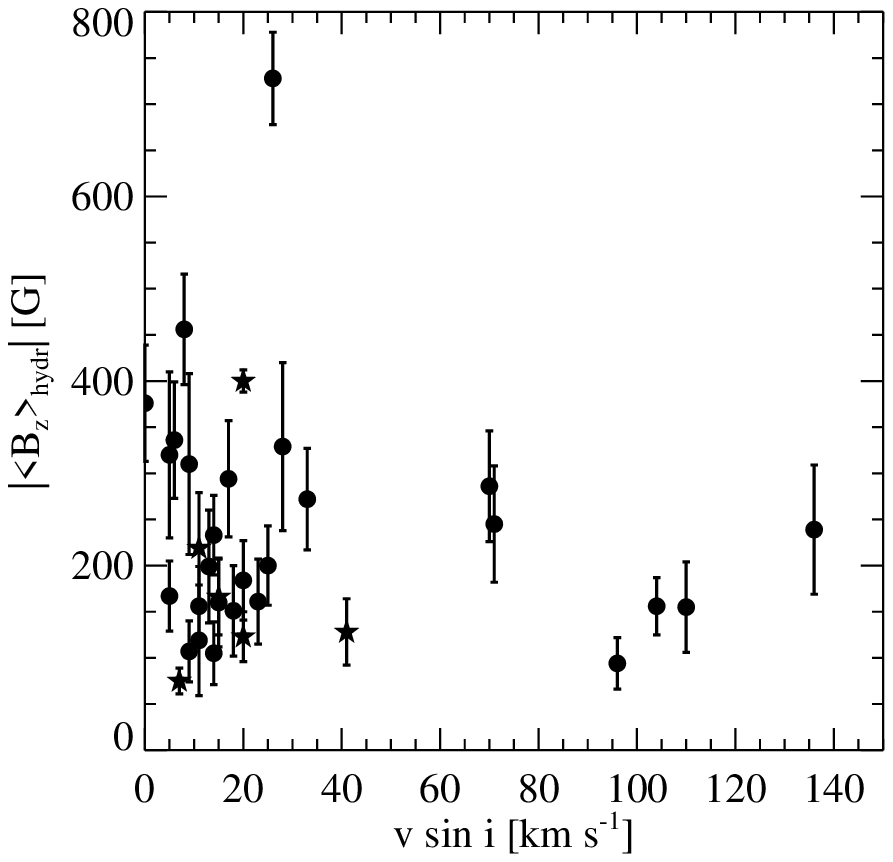}
\caption{
The strength of the longitudinal magnetic field measured with FORS\,1 using hydrogen lines
as a function of \vsini. The symbols are identical to those presented in Fig.~\ref{fig:f_Hhyd}.
}
\label{fig:vsini_Hhyd}
\end{figure}
%THIERRY
%There might be a way of grouping several of these figures (Figs.5-10) together.
%%%%%%%%%%%%

Out of 13 $\beta$\,Cephei stars 
studied to date with FORS\,1, four stars (31\%) possess weak magnetic fields, 
and out of the sample of six suspected $\beta$\,Cephei stars two stars show a weak magnetic field. 
The fraction of magnetic SPBs and candidate SPBs is found to be higher: roughly half of the 34 SPB stars (53\%)
were found 
to be magnetic and among the 16 candidate SPBs eight stars (50\%) possess magnetic fields.
In an attempt to understand why only a fraction of pulsating  stars exhibit magnetic fields, we
studied the position of magnetic and non-magnetic pulsating stars in the H-R diagram. 
Their distribution is shown in  Fig.~\ref{fig_peter}. In this figure 
filled circles correspond to confirmed SPB stars, open circles to 
candidate SPB stars, filled stars to confirmed $\beta$~Cephei stars, 
open stars to candidates $\beta$~Cephei stars, 
and squares to standard B stars. The stars with detected magnetic fields are presented 
by  symbols which are 1.5 times bigger than those for stars for which magnetic fields were not detected.
%the blue and red coloured symbols (stars and circles) correspond to the stars 
%with detected magnetic fields whereas filled symbols correspond to confirmed $\beta$\,Cephei and SPB 
%stars and open symbols to the candidate  $\beta$\,Cephei and SPB stars.
The instability strips shown in Fig.~\ref{fig_peter} were determined from
theoretical models for main-sequence stars with 2\,$M_\odot$ $\le M \le$ 15\,$M_\odot$.
For these models, the instability of the modes with  $\ell \le 3$ and
eigenfrequencies between 0.2 and 25\,d$^{-1}$ was checked. The hence derived instability
strips only contain models having unstable $\beta$~Cephei- and/or SPB-like
modes (De Cat et al.\ \cite{DeCat2007a}).
From the locations of the boundaries of the instability strips it is clear that
$\beta$~Cephei stars are not found
close to the ZAMS as they are not predicted to be there. Some concentration
of less massive SPB stars towards the
ZAMS is in agreement with expectations from evolutionary models, as the
evolution time near the TAMS
is faster than the evolution time close to the ZAMS. For hotter SPBs there
are only unstable modes in the stars close to the TAMS.
The magnetic SPB stars with masses $\leq$4$M_{\sun}$ seem to be concentrated closer to the ZAMS than the higher mass stars.
On the other hand, no clear picture emerges as to the possible evolution of the magnetic field across the 
main sequence.
%for the evolutionary age difference between SPBs and 
%candidate SPBs  with detected magnetic fields and stars with undetected magnetic fields.
 Among the most massive $\beta$\,Cep stars, two magnetic stars, $\xi^1$~CMa and HD\,50707, are located close to 
the boundary of the theoretical instability strip. $\delta$\,Cet  with the weakest 
magnetic field detected in our sample of $\beta$\,Cep stars is also the least massive target of this group and is 
located close to the center of the main sequence.
The 
%forth 
%THIERRY
 fourth
%%%%%%%%%%%%
 $\beta$\,Cep star with a detected magnetic field, HD\,180642, might be the youngest
object in our $\beta$\,Cep star sample, but the parameters are not very reliable (see Sect.~\ref{sect:sample}).
%THIERRY
%The parameters we derive for HD 180642 are based on extrapolation of the calibration grids and, as you mentioned, not very reliable. From spectroscopy, we get Teff~24 kK and logg somewhere between 3.4 and 3.8, so it is unlikely to be young. And then, as it is, it is not located in the instability strip, which is a bit of a problem (!). 
%%%%%%%%%%%%
From the position of magnetic and non-magnetic pulsating stars, including also the two normal 
B-type stars, HD\,153716 and the nitrogen rich star HD\,52089, it is obvious 
that their domains in the H-R diagram largely overlap. For this reason we suggest that the evolutionary age 
cannot be the decisive factor for the presence of a magnetic field in pulsating stars. 
%THIERRY
%The evolutionary age is not the main factor governing the evolution of the magnetic field in pulsating B stars, but what about the situation for lower mass stars? 
%If magnetic stars can be found anywhere along the MS, then that would mean that 1) the fields are fossil (very long decay times) or 2) dynamo processes are at work during the whole core-hydrogen burning phase.
%%%%%%%%%%%%
 This suggestion is also supported by 
%THIERRY
%the plot presented in 
%%%%%%%%%%%%
 Fig.~\ref{fig:f_Hhyd} where we present the strength of the longitudinal magnetic field 
measured in confirmed pulsating $\beta$\,Cep and SPB stars and candidate  $\beta$\,Cep and SPB stars as a 
function of the completed fraction of the main-sequence lifetime.
In this figure and all subsequent  figures, we use the data from this study
and from Hubrig et al.\ (\cite{Hubrig2006}).
The magnetic field strength is used in absolute values, without taking 
into account the polarity of the field.
No obvious trend for the change of the strength of the magnetic field across the H-R diagram can be detected.
%In Figs.~\ref{fig:M_Hhyd} and \ref{fig:teff_Hhyd} we present the strength of the magnetic field as a function
%of mass and effective temperature. 
We also find no trend between the distribution of the strength of the magnetic field
and stellar mass or effective temperature, though for the small sample of $\beta$~Cephei stars a 
slight increase of the strength of the magnetic field 
with the stellar mass and effective temperature is possible (Figs.~\ref{fig:M_Hhyd} and \ref{fig:teff_Hhyd}).
%THIERRY
%I would say it is tentative at best...
%%%%%%%%%%%%

In Table~\ref{table_freq} we present the frequencies and the corresponding pulsating amplitudes of all pulsating stars for 
which magnetic fields have been detected with FORS\,1 up to now.
For the candidate $\beta$~Cephei star HD\,136504 no amplitude is given
since only spectroscopic observations have been carried out. 
As we already mentioned in Sec.\,2, four magnetic $\beta$~Cephei stars, $\delta$\,Cet, $\xi^1$\,CMa, V2052\,Oph, 
and $\beta$\,Cep have another common property: they are either radial pulsators ($\xi^1$\,CMa)
or their pulsations are dominated by 
a radial mode ($\delta$\,Cet, $\beta$\,Cep, and V2052\,Oph).
In addition, Aerts (\cite{Aerts2000}) found HD\,180642
to be a large amplitude non-linear pulsator with a dominant radial mode.
Although the mode identification for the main frequency of the $\beta$~Cephei star HD\,50707 is not completely clear,
it has been shown by other authors that this star pulsates non-linearly (Shobbrook et al.\ \cite{Shobbrook2006}; Heynderickx \cite{Heynderickx1992})
as is the case for all magnetic $\beta$~Cephei stars for which the pulsational behaviour has been 
carefully studied in the past.
It is therefore quite possible that there might be a link between such a non-linear pulsation behaviour dominated by a 
radial mode and the presence of a magnetic field.

In Figs.~\ref{fig:freq_Hhyd} and \ref{fig:amp_Hhyd} we plot the strength of the magnetic field as a function of 
the pulsation frequency and the corresponding pulsating amplitudes.
For multiperiodic stars we use in Fig.~\ref{fig:freq_Hhyd} the frequency of the pulsation period with 
the highest amplitude. 
It is possible that stronger fields tend to be found in stars with lower pulsating frequencies and 
smaller pulsating amplitudes. 
%A somewhat similar trend is found if we consider a correlation
%between the field strength and \vsini{} values, i.e. 
%stronger magnetic fields tend to be found in more slowly rotating stars.
%Interestingly, it seems that stronger fields tend to be found in stars with 
%lower pulsation frequencies and 
% higher 
%THIERRY
 %lower
%%%%%%%%%%%%
 %pulsation amplitudes. 
%THIERRY
%These trends are completely biased by the star with the highest field value, and I don't see any evidence for correlations.
%%%%%%%%%%%%

In  Fig.~\ref{fig:vsini_Hhyd} we display the strength of the magnetic field as a function of \vsini{}-values.
The majority of pulsating stars have rather low \vsini{}-values, less than 30\,km/s, and it is possible that 
%slow rotating stars exhibit somewhat larger magnetic fields. 
%THIERRY
 the magnetic stars are rotating more slowly.
%%%%%%%%%%%%
Magnetic breaking and angular momentum transport along the field lines would offer a natural 
explanation for the slow rotation of our magnetic pulsating stars. 
Slow rotation is one of the main characteristics of Ap and Bp 
stars and it is generally assumed that Ap stars are slow rotators because of magnetic braking. 
However, this judgment about the slow rotation of pulsating stars is based on the assumption that 
the stars with the low \vsini{}-values are not actually viewed pole-on stars.
On the other hand, considering random inclination angles and the size of our sample,
the number of pole-on stars should be rather small. 
For no star is the rotation period known to date.
%THIERRY
%I'm not sure this argument about the orientation is completely valid. We have quite a large sample, so I guess it is unlikely to have a significant fraction of stars seen pole on just by chance (can be calculated assuming random inclinations of the lines of sight though...).  
%%%%%%%%%%%%
The rotation periods of pulsating stars can currently be rather easily determined by space-based monitoring with 
the up-coming mission of the nanosatellites BRITE (BRIght Target Explorer) or with the already in orbit microsatellite 
MOST (Microvariability and Oscillations of STars). 
These satellites can provide intense photometric monitoring to search 
and precisely identify pulsation modes and  rotationally split modes in both, SPB and  $\beta$\,Cep stars,
and in Bp stars (we refer to Sect.~\ref{sect:results} with the discussion related to He-strong stars). Such observations
will be crucial for the understanding of the 
generation mechanism of the magnetic field in hot B-type stars. 

In summary, we have demonstrated that a significant fraction of pulsating B stars are magnetic.
This is a very important result, which should be included in future discussions related to 
theoretical works similar to that of Hasan et al.\ (\cite{Hasan2005}).
The presented magnetic field measurements in pulsating stars confirm that their
longitudinal magnetic fields are rather weak in comparison
to the kG fields detected in magnetic Bp stars.
Although our results provide some new clues, the observational results presented in this work are still 
inconclusive as to the difference between magnetic and non-magnetic pulsating stars.
The present-day magnetic field data are far from sufficient to prove the existence of
either pulsational or rotational variability of magnetic fields in the studied stars. 
Additional future magnetic field measurements are also needed to study the indicated loose trends in the few
dependencies presented above.
%THIERRY
%Instead of focusing on these rather inconclusive results I would conclude on a positive note saying that you have demonstrated that a significant fraction of pulsating B stars are magnetic. This is a very important result and can be nicely discussed in relation with the (very few...) theoretical works done on this (e.g. Hasan et al. 2005).
%%%%%%%%%%%%
Abundance studies of nitrogen and boron in B-type stars are still scarce.
It will be important to carry out sophisticated abundance analyses for all
pulsating stars.
Besides confirming the use of the chemical anomaly as an indicator for the presence of magnetic fields,
it will also be of importance to obtain abundances 
for pulsating stars with non-detections.
Such analyses are necessary to convincingly reveal a dichotomy between the two groups.
%Such analyses are necessary to confirm the advantage of using the chemical 
%anomaly as an indicator for the presence of magnetic fields.
%THIERRY
%It is nice to see that the magnetic stars are N rich, but it would be much more convincing to show that the stars without detection tend to be N-normal. I've checked the literature for all the stars in Tables 1-3, and found CNO data only for HD 61068, HD 129557 and HD 215573. The first two are N rich and HD 215573 is normal. Perhaps HD 61068 and HD 129557 are also magnetic, but we only have one measurement for both... So my point is that it is important to get CNO abundances for both magnetic AND non magnetic stars if we want to convincingly reveal a dichotomy between the two groups. 
%%%%%%%%%%%%

\section*{Acknowledgments}
MB is Postdoctoral Fellow of the Fund for Scientific Research, Flanders.
%This research made use of the SIMBAD database at the Centre de Donn\'ees astronomiques de 
%Strasbourg (CDS), France.
We would like to thank to H.F. Henrichs for valuable comments. This research has made use of 
the SIMBAD database, operated at CDS, Strasbourg, France.

%\appendix

\label{lastpage}

\end{document}